\documentclass[12pt,reqno]{amsart}
\usepackage{amssymb, amsmath, hyperref}
\usepackage{pdfsync}
\usepackage{stmaryrd}
\usepackage{graphics,graphicx,fancyhdr,color}

\newtheorem{df}{Definition}[section]
\newtheorem{lm}[df]{Lemma}
\newtheorem{lemma}[df]{Lemma}
\newtheorem{prop}[df]{Proposition}
\newtheorem{thm}[df]{Theorem}

\makeatletter \@addtoreset{equation}{section}

\newcommand{\cal}{\mathcal}

\newcommand{\bes}{\begin{displaymath}}
\newcommand{\ees}{\end{displaymath}}
\newcommand{\be}{\begin{equation}}
\newcommand{\ee}{\end{equation}}
\newcommand{\ba}{\begin{eqnarray}}
\newcommand{\ea}{\end{eqnarray}}
\newcommand{\bas}{\begin{eqnarray*}}
\newcommand{\eas}{\end{eqnarray*}}
\newcommand{\@Bbb}[1]{\ensuremath{\mathbb #1}}

\newcommand{\B}{{\@Bbb B}}
\newcommand{\C}{{\@Bbb C}}

\newcommand{\T}{{\mathbb T}}
\newcommand{\F}{{\@Bbb F}}
\renewcommand{\P}{{\mathbb P}}
\newcommand{\bbP}{{\P}}
\newcommand{\bbE}{{\mathbb E}}
\newcommand{\Q}{{\@Bbb Q}}
\newcommand{\bQ}{{\@Bbb Q}}
\newcommand{\N}{{\@Bbb N}}
\newcommand{\R}{{\mathbb R}}

\newcommand{\bbR}{{\mathbb R}}
\newcommand{\W}{{\@Bbb W}}
\newcommand{\Z}{{\mathbb Z}}
\newcommand{\bbZ}{{\Z}}
\newcommand{\bbT}{{\T}}

\newcommand{\la}{\lambda}
\newcommand{\al}{\alpha}

\newcommand{\bt}{\beta}
\newcommand{\si}{\sigma}

\newcommand{\Om}{\Omega}
\newcommand{\om}{\omega}

\newcommand{\ep}{\varepsilon}
\newcommand{\eps}{\epsilon}

\newcommand{\@s}[1]{\ensuremath{\mathcal #1}}
\newcommand{\cA}{\@s A}
\newcommand{\cB}{\@s B}
\newcommand{\cC}{\@s C}
\newcommand{\cD}{\@s D}
\newcommand{\cE}{\@s E}
\newcommand{\cF}{\@s F}
\newcommand{\cG}{\@s G}
\newcommand{\cH}{\@s H}
\newcommand{\cI}{\@s I}
\newcommand{\cJ}{\@s J}

\newcommand{\cK}{\@s K}
\newcommand{\cL}{\@s L}
\newcommand{\cN}{\@s N}
\newcommand{\cM}{\@s M}
\newcommand{\cO}{\@s O}
\newcommand{\cP}{\@s P}
\newcommand{\cR}{\@s R}
\newcommand{\cS}{\@s S}
\newcommand{\cT}{\@s T}
\newcommand{\cV}{\@s V}
\newcommand{\cW}{\@s W}
\newcommand{\cX}{\@s X}
\newcommand{\cY}{\@s Y}
\newcommand{\cZ}{\@s Z}

\newcommand{\@bm}[1]{\ensuremath{\mathbf #1}}
\newcommand{\bma}{\@bm a}
\newcommand{\bmb}{\@bm b}
\newcommand{\bmc}{\@bm c}
\newcommand{\bmd}{\@bm d}

\newcommand{\bme}{\@bm e}
\newcommand{\bmf}{\@bm f}
\newcommand{\bmg}{\@bm g}
\newcommand{\bmh}{\@bm h}
\newcommand{\bmi}{\@bm i}
\newcommand{\bmj}{\@bm j}
\newcommand{\bmk}{\@bm k}
\newcommand{\bml}{\@bm l}
\newcommand{\bmm}{\@bm m}
\newcommand{\bmn}{\@bm n}
\newcommand{\bmo}{\@bm o}
\newcommand{\bmp}{\@bm p}
\newcommand{\bmq}{\@bm q}
\newcommand{\bmr}{\@bm r}
\newcommand{\bms}{\@bm s}
\newcommand{\bmt}{\@bm t}
\newcommand{\bmu}{\@bm u}
\newcommand{\bmw}{\@bm w}
\newcommand{\bmv}{\@bm v}
\newcommand{\bmx}{\@bm x}
\newcommand{\bx}{\@bm x}

\newcommand{\bmy}{\@bm y}
\newcommand{\bz}{\@bm z}

\newcommand{\by}{\@bm y}

\newcommand{\bmzero}{\@bm 0}

\newcommand{\ga}{\gamma}

\newcommand{\@g}[1]{\ensuremath{\mathfrak #1}}
\newcommand{\gA}{\@g A}
\newcommand{\gD}{\@g D}
\newcommand{\gJ}{\@g J}
\newcommand{\gF}{\@g F}
\newcommand{\gM}{\@g M}
\newcommand{\gR}{\@g R}

\newcommand{\commentout}[1]{{}}

\makeatother

\begin{document}

\title[Energy superdiffusion]{Superdiffusion of energy in a chain of harmonic
  oscillators with noise} 

\author{Milton Jara}
\address{Milton Jara\\ Institutio Nacional de matematica Pura e Aplicada\\Rio de Janeiro, Brazil.}
\email{mjara@impa.br}
\author{Tomasz Komorowski}
\address{Tomasz Komorowski\\Institute of Mathematics, Polish Academy Of Sciences\\Warsaw, Poland.}
\email{komorow@hektor.umcs.lublin.pl}
\author{Stefano Olla}
\address{Stefano Olla\\
 CEREMADE, UMR-CNRS 7534\\
 Universit\'{e} Paris Dauphine\\
 Paris, France.}
 \email{{\tt olla@ceremade.dauphine.fr}
}

 \begin{abstract}
{\em We consider a one dimensional infinite chain of harmonic oscillators whose
  dynamics is perturbed by a stochastic term conserving energy and
  momentum. We prove that in the unpinned case the macroscopic
  evolution of the energy converges to a fractional
  diffusion governed by $-|\Delta|^{3/4}$. For a pinned system we prove
  that its energy evolves diffusively, generalizing some  results of
  \cite{bo}.}  
\end{abstract}
\vspace*{-1in}
\date{\today. 
}
\thanks{
This paper has been partially supported by the
  European Advanced Grant {\em Macroscopic Laws and Dynamical Systems}
  (MALADY) (ERC AdG 246953), T. K. acknowledges the support of the
  Polish National Science Center grant DEC-2012/07/B/SR1/03320.}
\maketitle

\section{Introduction}

Superdiffusion of energy and the corresponding anomalous thermal
conductivity have been observed numerically in the dynamics of unpinned
FPU chains \cite{llp97,sll}. This is generally attributed to the small
scattering rate for low modes, due to momentum conservation. When the
interaction has a pinning potential, it is expected that the system
undergoes a
normal diffusion. 
  More recently the problem has been studied in models where the Hamiltonian
dynamics is perturbed by stochastic terms that conserve energy and
momentum, like random exchange of velocity between nearest neighbors
particles \cite{bborev,bbo2}. In these models, the interaction is
purely harmonic, and as a result, the Green-Kubo formula for thermal
conductivity $\kappa$ can be studied explicitly. It diverges for one and
two dimensional lattices in case no pinning potential is present, while thermal
conductivity stays finite for pinned systems or in dimension  $d\ge
3$. In the cases  
when the conductivity is finite it is proven in
\cite{bo} that energy fluctuations evolve diffusively following the heat
equation. 

The main result of the present article concerns the nature of the
superdiffusion in dimension 1, when the chain is unpinned. It has
already been proven that in the weak noise limit (where the average number of
\emph{stochastic collisions} is kept finite as in  {the} Grad limit) the
Wigner distribution converges to an inhomogeneous phonon {linear} Boltzmann
equation \cite{BOS}. Since the corresponding scattering kernel $R(k,k')$ is
positive, the resulting Boltzmann equation can be interpreted
probabilistically as the evolution of the density for some Markov
process: in this limit a \emph{phonon} of mode $k$ moves with the
velocity given by the gradient of the dispersion relation
$\nabla\omega(k)$ and change mode with rate $R(k,k')$. 
Under a proper space-time rescaling, this process converges
to a L\'evy superdiffusion generated by the fractional laplacian
$-|\Delta|^{3/4}$. This is proven in \cite{kjo,babo}, using
probabilistic techniques {such} as coupling and
martingale convergence theorems.  
A completely analytic proof of
the  convergence, from a kinetic  to a fractional
diffusion equation, without the use of the probabilistic
representation, has been proposed  in \cite{mmm}.  All these results
provide  a \emph{two-step} solution: 
first take a kinetic limit, then use a \emph{hydrodynamic} rescaling
of the kinetic equation. A kind of diagonal procedure is treated in
\cite{KS}: using the probabilistic approach one can push the time scale a little longer   (matching
suitably the size of a still small scattering rate) than in the kinetic
limit case. As a result it is possible to obtain  the diffusive limit
under the pinning potential and superdiffusive 
in the unpinned case. 

In the present paper we prove a direct limit to the fractional
superdiffusion, just by rescaling space and time, without the weak
noise assumption. 
We also recover the diffusive limit results of \cite{bo} in the case
of a finite diffusivity and  study the cases of intermediate weaker
noise limits. The rigorous formulations of our main  results are
listed in Section \ref{sec3a}.  

In a recent article \cite{Sp13}, Herbert Spohn predicts the same fractional
superdiffusive behavior for the heat mode in the $\beta$-FPU at zero
pressure. This follows  from an application of mode coupling
approximation procedure to fluctuating hydrodynamic equations.  {Our
  present results concern a model}, that has also three 
conserved quantities. They are  in agreement with  the predictions of
  \cite{Sp13}, confirming that the harmonic stochastic model is a good
  approximation of some non-linear models, at least in the case of symmetric
  interactions. 

The strategy of the proof is as follows: first,  we formulate the result
for the limit evolution of the Wigner distribution $W_\epsilon(t)$ of
the energy, when the initial data are in $L^2$, see Theorems
\ref{diffusive-1} and \ref{superdiffusive-1},  
proven in Sections
\ref{sec-homo} and \ref{sec-iden}. 
These results concern the system
with non--equilibrium initial data but of the finite total energy. 
 The  extension to homogeneous initial data  (whose
  $L^2$ norm  is infinite), in particular 
the equilibrium dynamics with Gibbs distributed
 initial data,  is possible
by a simple duality argument, see Section \ref{diff-scale-1}. 
 
Our results can be formulated in terms of a local energy functional,
see Theorems \ref{energy-prop-main} and \ref{thm-pinned} in the case
of the $L^2$ integrable initial data, and Theorem \ref{thm-fluct-heat} for the
initial data in equilibrium, respectively. This is possible thanks to
the asymptotic equivalence of the relevant energy functionals proven
in Propositions \ref{prop011404} and \ref{prop012204} below.


Concerning the proof when the initial data have square summable realizations, which is the
crucial part of our argument, we study first the time
evolution of the Wigner distribution of the energy $W_\epsilon(t)$,
that  represents the energy density in both the spatial variable and
 frequency modes (in fact, it is more convenient to work with
 the
  Fourier transform of $W_\epsilon(t)$ in the spatial variable). 
As
it has been already remarked in \cite{BOS}, the evolution of $W_\epsilon(t)$  is not autonomous but
involves another distribution $Y_\epsilon(t)$ whose real and imaginary parts represent the
difference between kinetic and potential 
energy and the energy current, respectively, see \eqref{exp-wigner-eqt-1ca}. The principal advantage of
working with the pair  $(W_\epsilon(t),Y_\epsilon(t))$ is that its
evolution can be described by a system of ordinary differential
equations. By performing the Laplace transform in the temporal domain
the system reduces further to an algebraic system of linear equations,
see \eqref{exp-wigner-eqt-1k}, and the problem of finding the
asymptotics of the energy density for the chain of oscillators reduces
to the question of asymptotics of solutions of the system. This
is done in Sections \ref{sec-homo} and \ref{sec-iden}.
First, we observe in Section \ref{sec-homo}, that  due to the high number of random collisions in the time scales
considered,  both $W_\epsilon(t)$ and $Y_\epsilon(t)$ homogenize
(unlike in  the case of the 
kinetic limit considered in \cite{BOS}), and their limits do not
depend on the frequency mode variable. The homogenization is proven in
Theorem \ref{cor011811a}. In addition, because of fast fluctuations, the
time integral of  
$Y_\epsilon(t)$ will disappear from the final equation, as in the case
of the kinetic limit in \cite{BOS}. The above implies that the
phonon-Boltzmann equation gives a good approximation of the evolution
of $W_\epsilon(t)$,  {but the presence of the error term, that is of
  order $o(1)$, as  
$\epsilon$ tends to $0$}, does not allow for a direct application of
the probabilistic 
approach of  \cite{kjo,babo}. Instead, we use a version of the
analytic approach of \cite{mmm}, based on projections on the product
components of the scattering kernel appearing in the homogenized
dynamics, see \eqref{060411} and \eqref{010301-15} below.  This is done in
Section \ref{sec-iden}.

 In our choice of the  dynamics  a diffusive random exchange of
 momenta takes
   place between the
three nearest neighbor particles in such a way that total kinetic energy and
 momentum are conserved in the process. 
 However, our method can be applied  to other linear models with quite general stochastic 
scattering mechanisms, generating different scattering rates.
The result does not depend on the particular type of stochastic
perturbations, as long as it  conserves the appropriate quantities.
E.g. we could consider 
a model  with   a simple Poissonian exchange  of  the two nearest
neighbor velocities described in Section \ref{sec-ex} below. 
%
In fact  this case is
  computationally less involved, due to a simpler structure of the
  respective scattering
kernel.

Concerning the possible generalizations of our results to dimensions $d\ge 2$,  see \cite{bborev}
  for the formulation of the model, we conjecture they    can also be treated by the
  present method. 

Although for the equilibrium fluctuations we prove only the
  convergence of the covariance function, our approach can be further developed
  to obtain the convergence in law for the equilibrium fluctuation
  field to the respective Ornstein-Uhlenbeck process.  
  The question of the convergence in probability for the non-equilibrium case
 could  possibly be more involved, as it requires the control of higher
  moments of the energy distribution. 

A remark  concerning  the
  initial data of the system is also in order. We choose the initial probability
  distributions  of  the velocities and 
  inter-particle distances whose energy spectrum satisfies condition
  \eqref{finite-energy1}.
This condition implies that the initial data are macroscopically centered (see Section
  \ref{sec:super}). 
  While this choice is quite natural in the situation of a
  pinned chain, it requires some explanation in the unpinned case. In
  the latter situation, if we start with non-centered initial
  conditions, their respective macroscopic averages will evolve, 
  at the hyperbolic space-time scale, following the linear wave
  equation. As a result,  they will disperse to infinity, since we start with
  the data whose realization has  a finite $L^2$ norm. This implies that at a larger 
  superdiffusive  time scale these averages will be null. 

We mention here the article \cite{bgj}, where a result
  similar to ours is proven, by very different
  techniques, for a dynamics with two conserved quantities (energy and
  volume) in the case when the initial data is given by a Gibbs
  equilibrium  measure.

\section{The dynamics}

\subsection{Infinite chain of interacting harmonic oscillators}

\subsubsection{Hamiltonian system}

The dynamics of the chain of oscillators  can be written formally  as  a Hamiltonian system of differential equations 
\begin{eqnarray}
&&\dot {\frak q}_{x}(t)=\partial_{\frak p_x}{\cal H}({\frak p}(t),{\frak q}(t))
\label{eq:bas}\\
&&\nonumber\\
&& \dot {\frak p}_x(t)=-\partial_{\frak q_x}{\cal H}({\frak p}(t),{\frak q}(t)),\quad x\in\bbZ.\nonumber
\end{eqnarray}
The formal Hamiltonian is given by  
\begin{equation}
\label{ham}
{\cal H}({\frak p},{\frak q}):=\frac12\sum_{x\in\bbZ}{\frak p}_x^2+\frac{1}{2}\sum_{x,x'\in\bbZ}\alpha_{x-x'}{\frak q}_x{\frak q}_{x'}
\end{equation}
and
we assume also (cf  \cite{BOS}) that:
 \begin{itemize}
 \item[a1)] $(\alpha_x)_{x\in\bbZ}$ is real valued and there exists
   $C>0$ such that $|\alpha_x|\le Ce^{-|x|/C}$ for all $x\in \bbZ$, 
  \item[a2)] $\hat\alpha(k)$ is also real valued 
and  $\hat\alpha(k)>0$ for $k\not=0$ and in case $\hat \alpha(0)=0$ we
have  $\hat\alpha''(0)>0$,
\item [a3)] to guarantee that the local energy functional, see \eqref{eq:energypin} below, is non-negative we
  assume that  $\alpha_{x}\le 0,$ $x\not=0$. 
 \end{itemize}
 Here $\hat \al(k)$ is the Fourier transform of sequence
 $(\al_x)_x$, defined as
 \begin{equation}
\label{032504}
  \hat \al (k)=\sum_x \al_x \exp\{-2\pi i x k\}, \quad k\in\bbT.
 \end{equation}
   The above conditions imply that both functions $x\mapsto\alpha_x$
   and $k\mapsto\hat\alpha(k)$ are even. In addition, $\hat\alpha\in
   C^{\infty}(\bbT)$.
 Define the {\em dispersion relation} as $\om(k):=\hat
 \al^{1/2}(k)$. In case when  $\hat\alpha(0)>0$ the dispersion
 relation belongs to
 $C^\infty(\bbT)$. If $\hat\alpha(0)=0$ we can write
\begin{equation}
\label{om2}
\om(k)=|\sin(\pi k)|\sqrt{\frac{\hat \al''(0)}{2\pi^2}}\varphi(\sin^2(\pi k)),
\end{equation}
where
$\varphi:[0,+\infty)\to(0,+\infty)$ is of $C^2$ class and
 such that $\varphi(0)=1$.

The ${\frak p}_x$ component stands for the velocity (or momentum, as the mass of each particle is taken equal
to 1) of the particle $x$.
In the pinned case,  $\hat\al(0) > 0$, the particle labelled with $x$ feels a
pinning harmonic potential centered at $ax$, where $a \ge 0$ is an arbitrary equilibrium
interparticle distance,
so  ${\frak q}_{x}$ should be
interpreted as the displacement of the position of the particle $x$
from the point $ax$. Since the dynamics is linear, it does not depend on
$a$, which assume is equal to $1$.

In the unpinned case, $\hat\al(0) = 0$, the system is translation
invariant, and only the interparticle distances are relevant for the
dynamics. So in the unpinned case the variables ${\frak q}_{x}$ are
defined up to a common additive constant. 
Therefore, the relevant quantities are  functionals of     the 
relative distances between the particles. An important example is 
a wave function defined in  Section \ref{sec5-1}. 
Its definition is unambiguous both in the pinned and unpinned cases.
In the unpinned case, the total momentum and the energy
of the chain are  formally conserved, (besides the volume of course). 
Since we insist on preserving these properties, we  choose a 
stochastic perturbation having the same  conservation laws. 
This can be done  either locally, via a time continuous stochastic
exchange of momentum considered in this paper, or
through a time discontinuous random exchange  of momentum mechanism
(see Section \ref{sec-ex}).  


\subsubsection{Continuous time noise}
\label{sec:cont-time-noise}


We add to the right hand side of \eqref{eq:bas} a local stochastic
term that conserves ${\frak p}_{x-1}^2+{\frak p}_x^2+{\frak
  p}_{x+1}^2$ and ${\frak p}_{x-1}+{\frak p}_x+{\frak
  p}_{x+1}$. The respective stochastic differential equations can be
written as
\begin{eqnarray}
&d{\frak q}_x(t) &={\frak p}_x(t)\; dt
\label{eq:bas1},\\
&&\nonumber\\
& d{\frak p}_x(t)&=\left[-(\alpha*{\frak
    q}(t))_x-\frac{\ga}{2}(\beta*{\frak p}(t))_x\right]dt \nonumber\\
&&\quad +\ga^{1/2}\sum_{z=-1,0,1}(Y_{x+z}{\frak p}_x(t))dw_{x+z}(t),\quad x\in\bbZ,\nonumber
\end{eqnarray}
with the parameter $\ga>0$ that determines the strength of the noise
in the system, and $(Y_x)$ are vector fields given by
\begin{equation}
\label{011210}
Y_x:=({\frak p}_x-{\frak p}_{x+1})\partial_{{\frak p}_{x-1}}+({\frak p}_{x+1}-{\frak p}_{x-1})\partial_{{\frak p}_{x}}+({\frak p}_{x-1}-{\frak p}_{x})\partial_{{\frak p}_{x+1}}.
\end{equation}
Here $(w_x(t))_{t\ge0}$, $x\in\bbZ$ are i.i.d. one dimensional, real
valued, standard Brownian motions,  that are  non-anticipative over
some filtered probability space 
$(\Om,{\cal F},\left({\cal F}_t\right),\bbP)$.  
Furthermore, 
 $\beta_x=\Delta\beta^{(0)}_x$, where   
$$
 \beta^{(0)}_x=\left\{
 \begin{array}{rl}
 -4,&x=0,\\
 -1,&x=\pm 1,\\
 0, &\mbox{ if otherwise.}
 \end{array}
 \right.
 $$
The lattice Laplacian of  $(g_x)_{x\in\bbZ}$ is defined as
  $\Delta g_x:=g_{x+1}+g_{x-1}-2g_x$. Let also $ \nabla g_x:=g_{x+1}-g_x$ and $ \nabla^*
g_x:=g_{x-1}-g_x$. 
For a future reference we let
$
\beta_{1,x}:=\nabla^*\beta^{(0)}_x.
$
  A simple calculation shows that
\begin{equation}
\label{beta}
\hat \beta(k)=8\frak s^2( k)\left[1+2\frak c^2(k)\right]=8\frak s^2(
k)+4\frak s^2(2 k)
\end{equation}
and
\begin{equation}
\label{beta-1}
\hat \beta_1(k)=(1-e^{-2i\pi k})
\left(4+e^{2\pi i k}+e^{-2\pi i k}\right),
\end{equation}
where, for the abbreviation sake, we have writtten
\begin{equation}
\label{021701}
\frak s(k):=\sin(\pi k)\quad \mbox{and}\quad  \frak c(k):=\cos(\pi
k),\quad k\in\bbT.
\end{equation}




\subsubsection{Random momentum exchange}

\label{sec-ex}

{Another possible stochastic dynamics that conserves
the volume,  energy and momentum (in the unpinned case) can be
obtained by a ''jump'' type mechanism of the momentum exchange.}
More precisely, let $( N_{x,x+1}(t))_{x\in\bbZ} ~~$ be
i.i.d.  Poisson processes with  intensity $3\ga/2$.
The dynamics of the position component $\left({\frak q}_x(t)\right)_{x\in\bbZ}$  is the same
as in \eqref{eq:bas1}, while the momentum 
$\left({\frak p}_x(t)\right)_{x\in\bbZ}$ is a c\`adl\`ag process given by
\begin{eqnarray}
\label{eq:bas1a}
d{\frak p}_x(t)=&& -(\alpha*{\frak q}(t))_xdt\nonumber\\
&&
\\
&&+\left[\nabla{\frak p}_x(t-)d N_{x,x+1}(t)+\nabla^*{\frak p}_x(t-)dN_{x-1,x}(t)\right],\quad x\in\bbZ.\nonumber
\end{eqnarray}

\section{Main results: Macroscopic evolution}

\label{sec3a}
\subsection{Remarks on hyperbolic scaling. Euler equations}

\label{sec:superd-behav-unpinn}

Consider now the unpinned case $\hat\alpha(0) =0$. For a configuration $({\frak p}_x(t), {\frak r}_x(t))_{x\in\bbZ}$ we define the energy per
atom:
\begin{equation}
  \label{eq:energy}
  \frak e_x(t) = \frac{\frak p_x^2(t)}2 - \frac 14 \sum_y \alpha_{x-y} (\frak
  q_x(t) - \frak q_y(t))^2.
\end{equation}
Thanks to condition a3) we have 
$  \frak e_x(t)\ge0$.  
Notice also that, since $\sum_x\al_x=0$, formally we have  $\sum_x \frak e_x(t) = \mathcal H({\frak p}(t),{\frak q}(t))$.

Define  $\frak r_x(t) = \frak q_x(t) - \frak q_{x-1}(t)$. Then $\sum_x \frak
r_x(t)$, when finite, represents the total \emph{length} of the system
when the equilibrium interparticle distance $a=0$.
The chain has three formally conserved  (also called balanced) quantities
\begin{equation}
\label{012304a}
  \begin{split}
    &\sum_x \frak r_x(t) \qquad \text{- volume (length),}\\
    &\sum_x \frak p_x(t) \qquad \text{- momentum,}\\
    &\sum_x \frak e_x(t) \qquad \text{- energy.}
  \end{split}
\end{equation}

Because the noise is added to the system, these are
  the 'only' conserved 
quantities. More precisely, the only stationary probability measures
for the  infinite dynamics
\eqref{eq:bas}, 
that are also translation invariant and have a finite density entropy property (see Definition 4.2.1 of \cite{bo14}), are mixtures of the Gibbs measures
\begin{equation*}
  d\mu_{T,\bar p,\tau} = \frac{1}{Z_{T, \bar p,\tau}}\exp\left\{-
    T^{-1} \left(\cal H - \bar p \sum_x \frak p_x - \tau \sum_x \frak
      r_x\right)\right\} \prod_x d{\frak r}_x d{\frak p}_x 
\end{equation*}
parametrized by the temperature $T$, momentum $\bar p$ and
tension $\tau$, properly defined locally by the
  appropriate DLR equations on their
conditional distributions (see Section 4 of \cite{bo14}).

It can be proven 
that after the hyperbolic space-time scaling, these
conserved quantities evolve deterministically following the system of
Euler equations:
\begin{equation}
  \label{eq:euler}
 \left\{
  \begin{split}
    \partial_t \bar r(t,y) &= \partial_y\bar p(t,y),\\
    \partial_t \bar p(t,y) &= \tau_1\partial_y\bar r(t,y), \\
    \partial_t \bar e(t,y) &= \tau_1\partial_y\left(\bar r(t,y)\bar
      p(t,y)\right),
  \end{split}
  \right.
\end{equation}
with the initial data 
$$
\bar
r(0,y)= \bar r_0(y),\quad \bar p(0,y)= \bar p_0(y), 
\quad\bar e(0,y) = \bar
e_0(y)
$$
determined by the limits of quantities given by \eqref{012304a} at
time $t=0$. Here the parameter $\tau_1$, called the sound speed, is defined  by 
\begin{equation}
\label{tau}
\tau_1:=\frac{\hat \al''(0)}{8\pi^2}.
\end{equation}
More precisely, consider the empirical distributions
associated to the conserved quantities:
$
\frak u_x (t):= \left(
    \frak r_x(t), \frak p_x(t), \frak e_x(t) \right) .
$
 Then, 
\begin{equation}
\label{020301-15}
  \lim_{\eps\to0+} \eps \sum_x J(\eps x) \frak u_x\left(\eps^{-1} t\right)
  = 
   \int_{\bbR} J(y) \bar u(t,y) \; dy, \qquad 
\end{equation}
with
 $J$ - a smooth test function with compact
support,
and the 
convergence holds in probability for any $t>0$, provided it holds
for the initial distribution at $t=0$. The  functions $\bar r_0,\bar
p_0,\bar e_0$ are assumed to belong to $C_0^\infty(\bbR)$ - the space of all smooth and
compactly supported functions.  The components of 
$\bar u(t,y) $$:= \left( \bar r(t,y), \bar p(t,y), \bar e(t,y)\right)$ 
satisfy \eqref{eq:euler}.
Note that  system \eqref{eq:euler}
decouples. Quantities $(\bar r(t,y),
\bar p(t,y))$ satisfy the linear wave equation. Define the energy of
the \emph{phonon modes as}
$$
\bar e_{\rm ph}(t,y):=\frac{\tau_1\bar r^2(t,y)}2 + \frac{\bar p^2(t,y)}2.
$$
The residual energy component, called the local
  temperature profile, is given by 
\begin{equation}
\label{loc-temp}
T(y):=\bar e_0(y)-\bar e_{\rm ph}(0,y).
\end{equation}
The above definition leads to the decomposition of the energy profile $\bar
e(t,y)$   
into the {\rm temperature profile}, that remains stationary under the
hyperbolic scaling, and  the {\em phononic energy} $\bar e_{\rm ph}(t,y)$
whose  evolution is driven by the
linear wave equation, see \eqref{eq:euler}.
Observe that, starting with compactly supported
initial data, the phonon energy will disperse to
infinity,  as $t\to \infty$, and the energy profile will converge (weakly) to the temperature
profile. This is the reason why at any larger time scale, we have
only to look at the evolution of the temperature profile.


In the case of a finite number of particles $N = [\eps^{-1}]$, with
periodic or other boundary conditions,
convergence in probability stated in \eqref{020301-15}  can be proven by using relative entropy methods, see \cite{OVY}
 and \cite{EO}. In fact in the latter paper the limit has been shown
 in the  
non-linear case, in the smooth regime of the Euler equations.
In the
infinite volume, starting with the initial distribution $\mu_\eps$ on
the space of  configurations $({\frak r}_x,{\frak
  p}_x)_{x\in\bbZ}$ satisfying
\begin{equation}
  \label{eq:l2b}
\sup_{\eps\in(0,1]}  \eps \; \langle \cal H({\frak p},{\frak q}) \rangle_{\mu_\eps} < +\infty,
\end{equation}
with $\langle \cdot\rangle_{\mu_{\eps}}$ denoting the expectation
with respect to $\mu_\eps$,
the relative entropy method cannot be applied.

The detailed analysis of the behavior of the energy component corresponding to the phononic modes, under the hyperbolic scaling is not the subject of the present paper and we shall deal with it in our future work.
Our purpose here is to go beyond the hyperbolic time scale and
understand the behavior of the energy component corresponding to the  local temperature profile on the diffusive or (if necessary) superdiffusive space-time
scale.

\subsection{Behavior of
the energy functional}
\label{sec3.2}

Our main results deal with the macroscopic behavior of
the energy functional, 
for a given configuration $({\frak p}(t),{\frak q}(t))$. The energy per site is defined as
\begin{equation}
  \label{eq:energypin}
  \frak e_x(t) := \frac{\frak p_x^2(t)}2 - \frac 14 \sum_y \alpha_{x-y} (\frak
  q_x (t)- \frak q_y(t))^2 + \frac{\hat\alpha(0)}{2} \frak q_x^2(t).
\end{equation}
In this section we shall assume that   condition 
\eqref{eq:l2b} is satisfied.
Denote by   $\bbE_\eps$ the expectation with respect to the product measure $\bbP_\eps:=\mu_{\eps}\otimes \bbP$.


\subsubsection{Superdiffusive behavior of the unpinned chain}

\label{sec:super}

We assume first that $\hat \alpha(0) =0$, i.e.  the pinning potential
vanishes and the Hamiltonian
dynamics conserves both the momentum and energy. 

Define, the energy spectrum of a configuration $({\frak p}_x,{\frak
  q}_x)_{x\in\bbZ}$ as 
$$
{\frak w}_\eps(k):=\langle |\hat{\frak p}(k)|^2+\hat \al(k) |\hat{\frak q}(k)|^2\rangle_{\mu_\eps}
,\quad k\in\bbT,
$$
where $\hat{\frak p}(k)$ and $\hat{\frak q}(k)$
are the Fourier transforms of $({\frak p}_x)$ and $({\frak
  q}_x)$, respectively (see Section
 \ref{sec-basic} below), and $\hat \al(k)$ is given by \eqref{032504}.
Assumption \eqref{eq:l2b} is equivalent with
\begin{equation}\label{finite-energy1}
    \sup_{\eps\in(0,1]}\eps\int_{\bbT}{\frak w}_\eps(k)dk<+\infty.
\end{equation}
In what follows we shall suppose a stronger integrability condition on
${\frak w}_\eps(k)$. Namely, we assume that
\begin{equation}\label{K-cond1}
    \sup_{\eps\in(0,1]}\eps^2\int_{\bbT}{\frak w}_\eps^2(k)dk<+\infty.
\end{equation}

According to the remark made below formula \eqref{eq:waveenergy} the above assumption implies that both 
\begin{equation}
\label{r1b}
\lim_{\eps\to0+}\eps\sum_x J(\eps x)\langle{\frak
  r}_x\rangle_{\mu_\eps}=0
\end{equation}
and
\begin{equation}
\label{p1b}
\lim_{\eps\to0+}\eps\sum_x J(\eps x)\langle{\frak
  p}_x\rangle_{\mu_\eps}=0, \quad \forall\,J\in C_0^\infty(\bbR).
\end{equation}
Suppose that the initial distribution of energy satisfies the
following assumptions:
\begin{equation}
\label{e1b}
\lim_{\eps\to0+}\eps\sum_x J(\eps x) \langle{\frak
  e}_x\rangle_{\mu_\eps} 
=\int_{\bbR}J(y) W_0(y)dy,
\end{equation}
where $W_0\in L^1(\bbR)$ (it is obviously non-negative).
\begin{thm}
\label{energy-prop-main}
Let $\delta = 3/2$, then, under the conditions on the initial
distribution stated in the foregoing, for any test function $J\in 
C^{\infty}_0([0,+\infty)\times \bbR)$ we  have: 
\begin{equation}
\label{041803}
 \lim_{\eps\to0+}\eps\sum_{x}\int_0^{+\infty} J(t,\eps x)  \bbE_{\eps}{\frak
   e}_x\left(\frac{t}{\eps^{\delta}}\right) dt=\int_0^{+\infty}\int_{\bbR}W(t,y) J(t,y)dtdy,
 \end{equation}
where $W(t,y)$ satisfies the fractional heat equation:
\begin{equation}
  \label{eq:frheat}
  \partial_t W(t,y) = - \hat c|\Delta_y|^{3/4}W(t,y)
\end{equation}
with the initial condition $W(0,y)=W_0(y)$
and 
\begin{equation}
\label{hatc-32a}
\hat c:=\frac{[\al''(0)]^{3/4}}{2^{9/4}(3\ga)^{1/2}}.
\end{equation}
\end{thm}


The proof of this result is a direct
  consequence of Theorem \ref{superdiffusive-1} and Proposition
  \ref{prop011404} formulated below.  
In fact, (as  can be seen from the aforementioned results)  it can be
formulated in a more general way to cover 
also the case of  a \emph{weaker}
  noise, i.e. parameter $\gamma$ can be replaced by $\eps^s\gamma_0$, 
 for some $s\in [0,1)$ and $\ga_0>0$. Then,  the result is still valid
 at the time scale corresponding to the exponent $\delta = (3-s)/2$. The limit $W(t,y)$
 is the same as in the case $s=0$,
covered by Theorem \ref{energy-prop-main}. 

%



\subsubsection{Diffusive behavior of the pinned chain}
\label{sec:diff-behav-pinn}

If $\hat \alpha(0) >0$ there is a pinning potential and the Hamiltonian
dynamics does not conserve the momentum. Energy is the only relevant
conserved quantity but it does not evolve at the hyperbolic space-time scale. 
Define
\begin{equation}
  \label{eq:sigma}
  \hat\sigma^2 := \int_\bbT \frac{[\omega'(k)]^2}{R(k)} dk,
\end{equation}
where 
\begin{equation}
\label{R-k}
R(k):=\frac{\hat\beta(k)}{4}.
\end{equation}
Since $\om'(k)\approx k$ and $R(k)\approx k^2$, as $k\ll1$ (see
\eqref{om2} and  \eqref{beta}), we have $\hat\sigma^2 < +\infty$
(it is infinite in the 
unpinned case, due to $\om'(k)\approx {\rm sign}\,k$).
 As a result, the evolution is diffusive
 and we have the following:

\begin{thm}\label{thm-pinned}
Let $s\in [0,1)$, $\gamma = \eps^{s}\gamma_0$.
Then, under the assumptions made in the foregoing, for any $J(t,y)$ as in Theorem \ref{energy-prop-main} we have
\begin{equation*}
 \lim_{\eps\to0+}\eps\int_0^{+\infty}\left[\sum_{x} J(t,\eps x)  \bbE_{\eps}{\frak
   e}_x\left(\frac{t}{\eps^{\delta}}\right) \right]dt=\int_0^{+\infty}\int_{\bbR}W(t,y) J(t,y)dtdy,
 \end{equation*}
with $\delta = 2-s$, where $W(t,y)$ satisfies the heat equation:
\begin{equation}
  \label{eq:heat}
  \partial_t W(t,y) = \hat c \partial_y^2 W(t,y).
\end{equation}
Here
\begin{equation}
\label{hatc1}
  \hat c = \frac{\hat \sigma^2}{\gamma_0} + 8 \gamma_0 \pi^2 \qquad
  \text{if}\ \  s =0,
\end{equation}
and
\begin{equation}
\label{hatc}
   \hat c = \frac{\hat \sigma^2}{\gamma_0} \qquad \text{if}\ \  0< s <1.
\end{equation}
\end{thm}

\bigskip

The above theorem follows directly from Theorem \ref{diffusive-1} and
the already mentioned Proposition
  \ref{prop011404} formulated below.

 \subsection{Equilibrium fluctuations}
 \label{sec:equil-fluct}

 The results formulated in Section \ref{sec3.2}  hold under the condition of finite
 microscopic total energy \eqref{eq:l2b}. By a duality argument they
 can be applied to obtain the following macroscopic behavior of the
 fluctuations when the system starts in an equilibrium measure
 $\mu_{\cal E_0, 0, 0}$.  
 For the fluctuations of the energy mode we assume that $\ga=\ga_0\eps^s$ for some $\ga_0>0$
 and $s\in[0,1)$. Consider the energy
 fluctuation field
 \begin{equation}
   \label{eq:eff}
   \tilde{\frak e}_\eps(t, J) = \sqrt \eps \sum_x J(\eps x) \left[\frak
   e_x\left(\frac{ t}{\eps^{\delta}}\right) - \cal E_0\right], \quad J\in
 C_0^\infty(\bbR),
 \end{equation}
 where $\frak
   e_x(t)$ is given by  \eqref{eq:energypin} and $\delta$ is
 chosen as before, i.e.  $\delta = (3-s)/2$ in the
 unpinned case, and $\delta = 2-s$ in the pinned one. The covariance
 field is defined as 
 \begin{equation}
   \label{eq:equi-cov}
   C^{(e)}_\eps(t, J_1, J_2) : = \mathbb E\left[ \tilde{\frak e}_\eps(t, J_1)
     \tilde{\frak e}_\eps(0, J_2) \right] ,\quad J_1,J_2
\in C_0^\infty(\bbR).
 \end{equation}
 The following theorem  is a direct corollary from Theorems
 \ref{diffusive-1a} and \ref{superdiffusive-1a}, and Proposition
 \ref{prop012204} formulated below.
 \begin{thm}\label{thm-fluct-heat}
For any functions $J_1,J_2\in C_0^\infty(\bbR)$ and $\phi\in
L^1[0,+\infty)$ we have   
\begin{equation*}
     \lim_{\eps\to 0} \int_0^{+\infty}\phi(t)C^{(e)}_\eps(t, J_1, J_2)dt =\int_0^{+\infty}\phi(t) C^{(e)}(t, J_1, J_2)dt
   \end{equation*}
where $C^{(e)}(t, J_1, J_2)$ satisfies the equation
\begin{equation*}
  \partial_t C^{(e)}(t, J_1, J_2) = C^{(e)}(t, \mathcal A J_1, J_2)
\end{equation*}
with the initial condition 
$$
C^{(e)}(0, J_1, J_2) = \cal E_0 \int_\bbR
J_1(y)J_2(y) dy,
$$ and $\mathcal A = -\hat c|\Delta_y|^{3/4}$ in the
unpinned, or  $\mathcal A = D\Delta_y$ in the pinned case,
respectively. Coefficients $\hat c$ and $D$ are the same as in
Theorems \ref{energy-prop-main} and \ref{thm-pinned}, respectively.
 \end{thm}

{\bf Remark.} We remark here that  Theorems \ref{energy-prop-main} through \ref{thm-fluct-heat}
hold also for the dynamics corresponding to the random momentum
exchange model
described by  \eqref{eq:bas1a}. 

\section{Some basic notation}

\label{sec-basic}

The one dimensional torus $\bbT$  is the interval
$[-1/2,1/2]$ with identified endpoints.
Let $\ell^2 $ be the space of all complex valued sequences
$(\psi_x)_{x\in\bbZ}$, 
equipped with the norm $\|\psi\|_{\ell^2}^2:=\sum_x|\psi_x|^2$.
For $m\in\bbR$ we introduce $h_m$  --  the space of sequences
$(\psi_x)_{x\in\bbZ}$, for which 
$\|\psi\|_{h_{m}}^2 :=\sum_x(1+x^2)^{m}|\psi_x|^2<+\infty$.

Given a sequence $(\psi_x)\in\ell^2$  define $ \hat
\psi:\bbT\to\mathbb C$ -
its Fourier transform - by \eqref{032504}.
Obviously $\hat \psi$ belongs to $L^2(\bbT)$ - the space of
all complex valued functions equipped with the norm 
$\|\hat \psi\|_{L^2(\bbT)}:=\langle \hat \psi,\hat \psi\rangle_{L^2(\bbT)}^{1/2}$, where
$$
\langle \hat \psi,\hat \phi\rangle_{L^2(\bbT)}:=\int_{\bbT}\hat
\psi(k)\hat \phi^*(k)dk. 
$$
Formula  \eqref{032504}
determines also an isometric isomorphism between $h_m$ and
$H^m(\bbT)$ - the
 completion of $C^\infty(\bbT)$  in  the norm 
  $
\|\hat \psi\|_{H^m(\bbT)}:=\|\psi\|_{h_{m}}. 
$
We have $H^0(\bbT) =L^2(\bbT)$.

For  an arbitrary  $J:\bbT\to\mathbb C$, $k\in\bbT$, $p\in\bbR$ and $\eps>0$ we define 
\begin{eqnarray}
\label{010906}
&&
\delta_{\eps}J (p,k):=\frac{1}{\eps}\left[J\left(k+\frac{\eps p}{2}\right)-J\left(k-\frac{\eps p}{2}\right)\right],\nonumber\\
&&
\\
&&
\bar J(k,p):=\frac{1}{2}\left[J\left(k+\frac{p}{2}\right)+J\left(k-\frac{ p}{2}\right)\right].\nonumber
\end{eqnarray}

Given a set $A$ and two functions $f,g:A\to\mathbb R_+$ we 
say that
$
f(x)\approx
g(x)$, $x\in A$ if there exists $C>1$ such that
$$
\frac{f(x)}{C}\le g(x)\le Cf(x),\quad \forall\,x\in A.
$$
We write
$g(x)\preceq f(x)$, when only the upper bound on $g$ is satisfied.

Denote by  ${\cal S}$ the set of  functions $J:\bbR\times
\bbT\to\mathbb C$ that are of $C^\infty$ class and such that for any
integers $l,m,n$ we have 
$$
\sup_{y\in\bbR,\,k\in\bbT} (1+y^2)^{n}|\partial_y^l\partial_k^mJ(y,k)|<+\infty.
$$ 
For  $J\in {\cal S}$ we let $\hat J$ be its Fourier transform in
the first variable, i.e.
$$
\hat J(p,k):=\int_{\bbR}e^{-2\pi i y p}J(y,k)dy.
$$

Let  $a\ge 1$. We introduce the norm
\begin{equation}
\label{norm-ta01}
\| J\|_{{\cal A}}:=\int_{\bbR}\sup_k |\hat J(p,k)|dp.
\end{equation}
By 
${\cal A}$
we denote the completions of ${\cal S}$ in the respective norm.

\subsection{Averaged Wigner transform}

For a given $\eps\in(0,1]$ we let  $\psi$ be a random element distributed on  $\ell^2$ according to
a Borel probability measure $\mu_\eps$. We assume that (cf \eqref{finite-energy1})
\begin{equation}
\label{psi}
K_1:=\sup_{\eps\in(0,1]} \int_\bbT dk\; \left[\eps\left<|\hat \psi(k)|^2
        \right>_{\mu_\eps}\right]^2 <+\infty,
\end{equation}
where $\langle\cdot\rangle_{\mu_\eps}$ is the expectation with respect
to $\mu_\eps$.


%
%
%
%
Define $W_\eps^{(0)},  Y_\eps^{(0)}\in {\cal A}'$ 
    \begin{equation}
 \label{wigner1}
\langle
W_\eps^{(0)},J\rangle:=\frac{\eps}{2}\int_{\bbR\times\bbT}\left\langle
  \hat \psi^*\left(k-\frac{\eps p}{2}\right)\hat
  \psi\left(k+\frac{\eps p}{2}\right)\right\rangle_{\mu_\eps} 
\hat  J^*(p,k)dpdk,
\end{equation}
and
\begin{equation}
 \label{wigner2}
\langle Y_\eps^{(0)},
J\rangle:=\frac{\eps}{2}\int_{\bbR\times\bbT}\left\langle \hat
  \psi\left(k+\frac{\eps p}{2}\right)\hat \psi\left(-k+\frac{\eps
      p}{2}\right)\right\rangle_{\mu_\eps} \hat J^*(p,k)dpdk
\end{equation}
for any $ J\in {\cal A}$. 
From the Cauchy-Schwartz inequality  we  get
$$
|\langle W_\eps^{(0)},J\rangle|\le \frac{\eps}{2}\|J\|_{\cal A}\left\langle\|\hat
  \psi\|_{L^2(\bbT)}^2\right\rangle_{\mu_\eps}.
$$
Thanks to Jensen's inequality we conclude from \eqref{psi}  that
 \begin{equation}\label{finite-energy}
     K_0:=\sup_{\eps\in(0,1]}\frac{\eps}{2}\left\langle\|\hat \psi\|_{L^2(\bbT)}^2\right\rangle_{\mu_\eps}<+\infty.
 \end{equation}
Therefore 
   \begin{equation}
 \label{A1}
\sup_{\eps\in(0,1]}(\| Y_\eps^{(0)}\|_{{\cal A}'}+\|
W_\eps^{(0)}\|_{{\cal A}'})\le 2K_0.
\end{equation}
Functional  $ W_\eps^{(0)}\in{\cal A}'$ is called the {\em averaged
  Wigner transform of}   $\psi$. We refer to $ Y_\eps^{(0)}$ as the {\em averaged
 anti-Wigner transform}.  By Plancherel's identity we obtain
\begin{equation}
\label{wigner-def}
\langle W_\eps^{(0)}, J\rangle
=\frac{\eps}{2}\sum_{x,x'\in\bbZ}\left\langle \left( \psi_{x'}\right)^*
  \psi_x\right\rangle_{\eps}\int_{\bbT}e^{2\pi i(x'-x)k}
J^*\left(\frac{\eps}{2}(x+x'),k\right)dk, 
\end{equation}
 for any $J\in {\cal S}$. 
As a consequence of \eqref{A1},  both
  $\left( W_\eps^{(0)}\right)_\eps$ and  $\left( Y_\eps^{(0)}\right)_\eps$ are $*-$weakly (sequentially) compact in ${\cal A}'$, as $\eps\to0+$, i.e.
for any sequence $\eps_n\to0$ we can choose a subsequence  $(
W_{\eps_n'}^{(0)},  Y_{\eps_n'}^{(0)})_{n\ge1}$ whose each component
is $*-$weakly convergent 
in ${\cal A}'$, see Section 4.1 of \cite{BOS}.

One can show, see Theorem B4 of \cite{LS}, that if
$(
W_{\eps_n'}^{(0)})_n$ is $*-$weakly convergent
then there exists a finite Borel measure $W_0(dy,dk)$ on
$\bbR\times\bbT$ whose total mass does not exceed $K_0$ and such that
$$
\lim_{n\to+\infty}\langle  W_{\eps_n}^{(0)},
J\rangle=\int_{\bbR\times\bbT}  J^*(y,k)W_0(dy,dk),\quad J\in {\cal A}.
$$
Applied to functions $J(y,k) = J(y)$ the Wigner
distribution becomes:
\begin{equation}
  \label{eq:waveenergy}
  \langle W_\eps^{(0)}, J \rangle = \frac{\eps}2 \sum_x \langle
  |\psi_x|^2\rangle_{\mu_{\eps}} J(\eps x).
\end{equation}
{\bf Remark.} Observe that condition \eqref{psi} implies that $(\psi_x)_x$ is
centered in the following sense: for any $J\in C_0^\infty(\bbR)$ we have
\begin{equation}
  \label{eq:centered}
 \lim_{\eps\to 0+} \eps \sum_x \langle
  \psi_x\rangle_{\mu_{\eps}} J(\eps x)=0.
\end{equation}
Indeed, by Plancherel's identity we can write that the
absolute value of the expression under the limit equals
\begin{equation}
  \label{eq:centered1}
\eps\left| \int_{\bbT} \langle
  \hat\psi(k)\rangle_{\mu_{\eps}} \hat J_\eps(k)dk\right|,
\end{equation}
where
$$
\hat J_\eps(k):=\sum_x J(\eps x)\exp\left\{-2\pi i k x\right\}\approx \frac{1}{\eps}\hat J\left(\frac{k}{\eps}\right)
$$
and $\hat J(k)$ is the Fourier transform of $J(x)$. Expression in
\eqref{eq:centered1} is therefore estimated as follows
$$
\left|\int_{\bbT} \langle
  \hat\psi(k)\rangle_{\mu_{\eps}} \hat
  J\left(\frac{k}{\eps}\right)dk\right|\le \left[\int_{\bbT} dk\langle
  |\hat\psi(k)|^2\rangle_{\mu_{\eps}}^2 \right]^{1/4}\left[\int_{\bbT} \left|\hat
  J\left(\frac{k}{\eps}\right)\right|^{4/3}dk\right]^{3/4},
$$
where the estimate follows by H\"older inequality. Using
the change of variables $k':=k/\eps$ in the second integral on the
right hand side we conclude that it is bounded by
$\eps K_1^{1/4}\|\hat J\|_{L^{4/3}(\bbR)}$ for $\eps\in(0,1]$, which
proves \eqref{eq:centered}.


\subsection{Homogeneous random fields on $\bbZ$}

Suppose that ${\cal E}:\bbT\to[0,+\infty)$ is a Borel measurable function such that
 \begin{equation}
 \label{011210a}
 \sum_{x\in\bbZ}\left|\int_{\bbT} {\cal E}(k) e^{ 2\pi i k x} dk\right|<+\infty.
 \end{equation}
Let  $(\xi_y)_{y\in\bbZ}$ be a sequence of  i.i.d. complex Gaussian
random variables such that $\bbE\xi_0=0$ and $\bbE|\xi_0|^2=1$. Define
\begin{equation}
\label{053110}
\hat\psi(k)=\sum_{x\in\bbZ}\xi_x {\cal E}^{1/2}(k)e^{- 2\pi ik x}
\end{equation}
a  Gaussian, random $H^{-m}(\bbT)$-valued element, where $m>1/2$. 
 Its covariance  field equals
\begin{equation}
\label{021310aa}
{\cal C}(J_1,J_2):=\bbE\left[\langle J_1,\hat\psi\rangle\langle J_2,\hat\psi\rangle^* \right]=\int_{\bbT}{\cal E}(k)J_1(k)J_2^*(k)dk
\end{equation}
for any $J_1,J_2\in C^\infty(\bbT)$. Then,
$$
\psi_x:=\int_{\bbT}e^{2\pi i k x} \hat \psi(k)dk, \quad x\in\bbZ,
$$
is
a complex Gaussian, stationary field.
Function ${\cal E}(k)$ is  called  the {\em spectral measure} of the
field $(\psi_x)_{x\in\bbZ}
$.
In the particular case when ${\cal E}(k)\equiv 2{\cal E}_0$ we 
denote by $\mu_{{\cal E}_0}$ the
law of the  respective  field $(\psi_x) _{x\in\bbZ}$.
It is
 supported in 
$
h_{-m}
$, if  $m>1/2$. One can verify that
 $(\psi_x)_{x\in\bbZ}$ satisfies
\begin{equation}
\label{beta-temp}
\langle \psi_x\rangle_{\mu_{{\cal E}_0}}=0,\quad \langle \psi_x
\psi_{x'}\rangle_{\mu_{{\cal E}_0}}=0, \quad \langle \psi_x^*
\psi_{x'}\rangle_{\mu_{{\cal E}_0}}=2{\cal E}_0\delta_{x,x'},\quad  x,x'\in\bbZ.
\end{equation}

\section{Finite Macroscopic Energy: initial data in $L^2$}

\label{sec5.0}


 \subsection{The wave function and its evolution}
\label{sec5-1}

The wave function, adjusted to the macroscopic time, is defined as
(see \cite{BOS})
\begin{equation}
\label{011307}
\psi^{(\eps)}_x(t):= \left(\tilde{\om} * {\frak
  q}\left(\eps^{-\delta} t\right)\right)_x 
+i{\frak p}_x\left(\eps^{-\delta} t\right),\quad x\in\bbZ,
\end{equation}
where  $({\frak p}_x(t),{\frak q}_x(t))_{x\in\bbZ}$ satisfies \eqref{eq:bas1} and $\delta\in[0,2]$ is to be chosen later. 
Function $(\tilde \om_x)_{x\in\bbZ}$ is the inverse Fourier transform of the  dispersion relation function  $\om(k):=\sqrt{\hat \alpha (k)}$. The Fourier transform of the wave function is given by
\begin{equation}
\label{011307a}
\hat\psi^{(\eps)}(t,k)=\om(k)\hat {\frak q}\left(\frac{t}{\eps^{\delta}},k\right)+i\hat{\frak p}\left(\frac{t}{\eps^{\delta}},k\right),\quad k\in\bbT.
\end{equation}
Since ${\frak p}_x(t),{\frak q}_x(t)$ are real valued we have
\begin{equation}
\label{011307aa}
(\hat\psi^{(\eps)})^*(t,-k)=\om(k)\hat {\frak q}\left(\frac{t}{\eps^{\delta}},k\right)-i\hat{\frak p}\left(\frac{t}{\eps^{\delta}},k\right).
\end{equation}
From  \eqref{eq:bas} we conclude that
$\left(\hat\psi^{(\eps)}(t)\right)_{t\ge0}$ is an $L^2(\bbT)$ - valued,
adapted process that is the unique  solution of the It\^o stochastic
differential equation, understood in the mild sense (see e.g. Theorem  7.4 of \cite{daza})
  \begin{eqnarray}
 \label{basic:sde:2}
 &&
 d\hat\psi^{(\eps)}(t,k)=\left\{\frac{-i\om(k)}{\eps^{\delta}}\hat\psi^{(\eps)}(t,k)-\frac{\ga
   R(k)}{\eps^{\delta}}\left[\hat\psi^{(\eps)}(t,k)-(\hat\psi^{(\eps)})^*(t,-k)\right]\right\}dt\nonumber\\
 &&
 \\
 &&
  +\frac{i\ga^{1/2}}{\eps^{\delta/2}}\int_{\bbT}
  r(k,k')\left[\hat\psi^{(\eps)}(t,k-k')-(\hat\psi^{(\eps)})^*(t,k'-k)\right]B(dt,dk'),\nonumber
 \end{eqnarray}
 where $ \hat\psi^{(\eps)}(0)\in L^2(\bbT)$,  
$R(k)=\hat\beta(k)/4$,   and
 \begin{equation}
 \label{r}
 r(k,k'):= 
 2 \frak s^2(k)\frak s(2(k-k'))+2 \frak s(2k)\frak s^2(k-k')
 ,\quad k,k'\in \bbT.
 \end{equation}
The process $B(dt,dk)$ is a cylindrical Wiener noise on $L^2(\bbT)$
given by
$$
B(dt,dk)=\sum_{x\in\bbZ}w_x(dt)e_x^*(k) dk,
$$
where $(w_x)$ are i.i.d. standard, $1$-dimensional real Brownian
motions.

\subsection{Asymptotics of the Wigner transform}

In what follows we assume that  condition \eqref{psi} holds. Suppose also that $s\in[0,1)$ 
and that $\gamma=\ga_0\eps^{s}$. The noise  in \eqref{eq:bas1} is called weak (resp. strong) if $s>0$ (resp. $s=0$).
Furthermore assume that for any $J\in {\cal S}$ such that
$J(y,k)\equiv J(y)$ we have
\begin{equation}
\label{star-conv}
\lim_{\eps\to0+} W_\eps^{(0)}(J)= \int_{\bbR}W_0(y)J(y)dy,
\end{equation}
where $W_0(\cdot)$ belongs to
 $ L^1(\bbR)$ and is  non-negative.
  Its  Fourier transform shall be denoted by 
\begin{equation}
\label{W0}
\overline{W}_0(p):=
 \int_{\bbR}e^{-2\pi i p y }W_0(y)dy.
 \end{equation}

Since the total energy of the system $\sum_{x\in\bbZ}|\psi_x(t)|^2$ is conserved in time, see Section 2 of \cite{BOS}, 
for each $\eps\in(0,1]$ we have
\begin{equation}
\label{psi-as}
\|\hat \psi^{(\eps)}(t)\|_{L^2(\bbT)}=\|\hat \psi\|_{L^2(\bbT)},\qquad t\ge0,\quad \bbP_\eps \mbox{ a.s.}
\end{equation}
Let $ W_{\eps}(t)$ be the (averaged) Wigner transform of $\psi^{(\eps)}(t)$
given by
   \begin{equation}
 \label{wigner-t}
\langle W_\eps(t),J\rangle:=\frac{\eps}{2} \int_{\bbR\times\bbT}
\bbE_{\eps} \left[(\hat \psi^{(\eps)})^*\left(t,k-\frac{\eps p}{2}\right) 
 \hat \psi^{(\eps)}\left(t, k+\frac{\eps p}{2}\right)\right] 
\hat  J^*(p,k) dp dk.
\end{equation}
Here, as we recall,   $\bbE_\eps$ is the expectation with respect to 
$\bbP_\eps=\mu_{\eps}\otimes \bbP$. 
From \eqref{psi-as}
we
conclude, thanks to \eqref{A1},  that
\begin{equation}
\label{W-Y}
\sup_{\eps\in(0,1]}\sup_{t\ge0}\| W_{\eps}(t)\|_{{\cal A}'}\le K_0,
\end{equation}
where $K_0$ is the constant appearing in condition \eqref{finite-energy}.
As a direct consequence of the above estimate we infer that the family $\left(W_\eps(\cdot)\right)_{\eps\in(0,1]}$ is 
 $*-$weakly sequentially compact in any
 $L^\infty([0,T];{\cal A}')$, where $T>0$.

 Our main result states that, given $s\in[0,1)$, the
exponent  $\delta$ can be adjusted so that $\left(W_{\eps}(\cdot)\right)$ is $\vphantom{1}^*$-weakly
convergent, as $\eps\to0+$,   in any
 $L^\infty([0,T];{\cal A}')$, where $T>0$. The cases of  pinned ($\hat \al(0)>0$) and  unpinned chains  ($\hat\al(0)=0$) are 
 considered in Sections \ref{sec5.2.1} and  \ref{sec5.2.2} respectively.
 Before presenting our results  let us recall briefly the case of the kinetic limit treated in
 \cite{BOS}, see Theorem 5 in ibid.,  corresponding to
$s=1$, which is outside of  the scope of our results. 
Then, taking $\delta=1$ the family  $W_{\eps}(\cdot)$ is $\vphantom{1}^*$-weakly convergent, as $\eps\to0+$, to 
the unique weak solution of the linear kinetic equation
 \begin{equation}
\label{lattice-wave}
\partial_t W(t,y,k)+\frac{\om'(k)}{2\pi}\partial_yW(t,y,k)
 =\ga_0{\cal L}W(t,y,k).
\end{equation}
 The
scattering operator  ${\cal L}$, acting on the $k$-variable, is  defined 
by
 \begin{equation}
\label{scat}
 {\cal L} w(k):=2\int_{\bbT} R(k,k') w(k') dk'-2R(k)w(k),\quad w\in L^1( \bbT).
 \end{equation}
Here $R(k)$ is given by \eqref{R-k} and 
 \begin{equation}
 \label{060411}
R(k,k')
:=
\frac{3}{4}\sum_{\iota\in\{-,+\}} \frak e_\iota(k)  \frak
e_{-\iota}(k'),
\end{equation}
with 
\begin{equation}
\label{frak-e}
\frak{e}_+(k):=\frac{8}{3}{\frak s}^4( k),\quad
\frak{e}_{-}(k):=2{\frak s}^2(2 k).
\end{equation}
Note that 
\begin{equation}
\label{032110}
R(k)=\int_{\bbT}R(k,k')dk'=\frac{3}{4}\sum_{\iota\in\{-,+\}} \frak e_\iota(k) . 
 \end{equation}
From \eqref{beta} we conclude that 
 \begin{equation}
\label{011901-15}
R (k)\approx\sin^2(\pi k),\quad k\in \bbT.
\end{equation}




%


\subsubsection{Case of a pinning potential - diffusive transport of energy}

\label{sec5.2.1}

Suppose that
\begin{equation}
\label{al}
\hat \al(0)=\sum_x\al_x>0.
\end{equation}
 Since  \eqref{al} together with the assumption $\hat\al''(0)>0$ imply that 
$$
|\om'(k)|\approx |\sin(\pi k)|,\quad k\in \bbT.
$$
From the above and \eqref{011901-15} we infer that $\hat\si^2$ given
by 
formula  \eqref{eq:sigma} is finite.
\begin{thm}
 \label{diffusive-1}
 Assume that conditions  \eqref{star-conv} and \eqref{al} are in force and $\delta=2-s$, where $s\in[0,1)$. Then, for any $T>0$ the Wigner transforms
 $ W_\eps(\cdot)$ converge, as $\eps\to0+$, in the $*-$weak sense in 
   $L^\infty([0,T];{\cal A}')$ to $ W(\cdot)$ given by
\begin{equation}
\label{W-F}
W(t,y):=\int_{\bbR} e^{2\pi i p y}  \widehat W(t,p)dp,
\end{equation}
where, 
\begin{equation}
\label{heat1c}
 \widehat W(t,p)=\exp\left\{-\frac{\hat c p^2 t}{2}\right\}\overline W_0(p),\qquad
t\ge 0
\end{equation}
and  $\hat c$ is defined  by  \eqref{hatc},
if $s\in(0,1)$ (weak noise), or 
 by \eqref{hatc1}, 
 if   $s=0$ (strong noise).
 \end{thm}







\subsubsection{Case of a no pinning potential   - $3/2$ fractional superdiffusion}

\label{sec5.2.2}

Suppose that
\begin{equation}
\label{al-1}
\hat \al(0)=\sum_x\al_x=0.
\end{equation}
Recall that in this case 
the dispersion relation satisfies \eqref{om2}.
Therefore, the integral appearing on the right hand side of
\eqref{eq:sigma} becomes divergent.
Define  
\begin{equation}
\label{superheat1c}
 \widehat W(t,p)=\exp\left\{-\hat c |p|^{3/2} t\right\}\overline W_0(p),\qquad
t\ge0,
\end{equation}
with $\overline W_0(p)$ given by \eqref{W0} and
\begin{equation}
\label{hatc-32}
\hat c:=\frac{[\al''(0)]^{3/4}}{2^{9/4}(3\ga_0)^{1/2}}.
\end{equation}
Our result can be formulated as follows.
\begin{thm}
 \label{superdiffusive-1}
 Assume that  \eqref{star-conv} and \eqref{al-1} are in force. Then, the convergence assertion made in Theorem \ref{diffusive-1} still holds
 for any
  $s\in[0,1)$ and $\delta=(3-s)/2$. The limit  $ W(t)
 $    is
 given by \eqref{W-F} and 
 \eqref{superheat1c}. 
 \end{thm}
The proofs of the above two theorems are presented in Section \ref{diff-scale}.




\subsection{Energy modes}

\label{sec3.3}

Thanks to condition  $\frak e_x(t)$ defined in \eqref{eq:energypin}
are non-negative.
A simple calculation, using the definition of the Wigner transform, see \eqref{wigner-def}, shows  that
\begin{equation}
\label{011404}
\frac\eps2\sum_xJ(\eps x)\bbE_{\eps}|\psi_x^{(\eps)}(t)|^2=\langle
W_\eps(t),J\rangle, \quad J\in C_0^\infty(\bbR).
\end{equation}
 Theorem
 \ref{energy-prop-main} (resp. Theorem \ref{thm-pinned})
is a consequence of  \eqref{011404},  Theorem  \ref{superdiffusive-1}
(resp. Theorem \ref{diffusive-1})
and the following result, proved in Section \ref{sec:equiv-betw-wign}.
\begin{prop}
\label{prop011404} Suppose that $\hat \al(0)>0$ (resp. $\hat
\al(0)=0$) and 
that condition \eqref{psi} holds.
  Then,
\begin{equation}
\label{011803}
\lim_{\eps\to0+}\eps\sum_{x}J(\eps x)  \bbE_{\eps}\left[{\frak
    e}_x\left(\frac{t}{\eps^{\delta}}\right)
  -\frac12\left|\psi_x^{(\eps)}\left(t\right)\right|^2\right]=0,\quad t\ge0,\,J\in C_0^\infty(\bbR)
\end{equation}
for $\delta$ as in the
statement of Theorem \ref{diffusive-1} (resp. Theorem \ref{superdiffusive-1}).
\end{prop}







\section{Fluctuations in equilibrium}

\label{sec6.0}

In this section we assume that the system is in equilibrium, i.e. that
$(\psi_x)_{x\in\bbZ}$ is a homogeneous, complex Gaussian random field whose
covariance function is given by \eqref{beta-temp}. As we have already
mentioned, its law  $\mu_{{\cal E}_0}$ is supported in 
$h_{-m}$  for $m>1/2$  and the Fourier transform $\hat \psi(k)$
belongs to $H^{-m}(\bbT)$, $\mu_{{\cal E}_0}$ a.s.

Define $(\psi_x^{(\eps)}(t))$ as the field given by the
Fourier coefficients of 
the solution $(\hat\psi^{(\eps)}(t,k))$ of the equation
\eqref{basic:sde:2} whose  initial data is distributed according to $\mu_{{\cal E}_0}$. It has been shown in \cite{KOR},   see 
Proposition 2.1,  that  there exists a unique solution of the
equation,   understood in the mild sense  in
$C([0,+\infty);H^{-m}(\bbT))$  for any $m>1/2$, in case $\om(0)>0$ and
for $m\in(1/2,3/2)$, in the unpinned case. Furthermore, see Section 5.1
of ibid., the law  of $\hat \psi$ in $H^{-m}(\bbT)$ is  invariant in time under
the dynamics  determined by \eqref{basic:sde:2}.

 \subsection{Fluctuating Wigner distribution}

For a given $J\in{\cal S}$ define  the random Wigner transform as the field
\begin{equation}
\label{wigner-def1}
 {\widetilde{\cal W}}_\eps(\psi;J):= 
\frac{\sqrt\eps}{2} \sum_{x,x'\in\bbZ} \left(
   \psi_{x'}^* \psi_x -2\delta_{x,x'}  \mathcal
   E_0\right)\tilde J^*\left(\frac{\eps}{2}(x+x'),x'-x\right), 
\end{equation}
where
$$
\tilde J(y,x) = \int_{\bbT} e^{2\pi i xk} J\left(y,k\right)
dk,\quad (y,x)\in\bbR\times\bbZ.
$$ 
We will also denote ${\widetilde{\cal W}}_\eps(t;J) :=
 {\widetilde{\cal W}}_\eps(\psi^{(\eps)}(t);J)$. 
From the  time  invariance of the law of $\psi^{(\eps)}(t)$ and \eqref{beta-temp} we obtain
 $$
\bbE\widetilde{\cal W}_\eps(t;J)\equiv  \left\langle \widetilde{\cal
    W}_\eps(0;J) \right\rangle_{{\cal E}_0}= 0,\quad t\ge0.
 $$

 Given $J_1,J_2\in{\cal S}$ define also the covariance field
  \begin{equation}
\label{wigner-cov}
C_\eps(t;J_1,J_2):= \bbE\left[\widetilde{\cal W}_\eps(t;J_1)
  \widetilde{\cal W}_\eps(0;J_2)^*\right].
\end{equation}
{In the particular case when $t=0$ we obtain
\begin{eqnarray}
\label{covariance}
    &&C_\eps(0;J_1,J_2)= \left\langle\widetilde{\cal W}_{\eps}(0;J_1)
      \widetilde {\cal W}_\eps(0;J_2)^*\right\rangle_{{\cal E}_0} \\
    &&= \cal E_0^2 \eps\sum_{x,x'} \tilde J_1^*\left( \frac{\eps(x+x')}{2}, x'-x\right)
   \tilde J_2\left( \frac{\eps(x+x')}{2}, x'-x\right).\nonumber
\end{eqnarray}
Using the Parseval identity we conclude that 
\begin{equation}
\label{030201}
\lim_{\eps\to0+}C_\eps(0;J_1,J_2)= 
 {\cal E}_0^2 \int_{\bbR\times\bbT}    J^*_1(y,k) J_2(y,k)  dydk,
\end{equation}
for any $J_1,J_2\in{\cal S}$ and in the case  $J_m(y,k) \equiv
J_m(y)$, $m=1,2$ we
have 
\begin{equation}
\label{030201a}
\lim_{\eps\to0+}C_\eps(0;J_1,J_2)= 
 {\cal E}_0^2 \int_{\bbR}    J_1(y) \tilde J^*_2(y)  dy.
\end{equation}}

\subsection{Statements of the results}

\subsubsection{Case of a pinning potential}

Recall, see \eqref{W0}, that 
$$
\bar J(p):=\int_{\bbT}   \widehat J(p,k)dk,\quad J\in{\cal S}.
$$
Our result dealing with this situation can be formulated as follows.
\begin{thm}
 \label{diffusive-1a}
If $\hat \al(0)>0$ and $\delta=2-s$, where $s\in[0,1)$ then,
\begin{equation}
 \label{wigner-evolve-1}
 \begin{split}
&\lim_{\eps\to0+}\int_0^{+\infty}\phi(t)C_\eps(t;J_1,J_2)dt\\
&= {\cal E}_0^2
\int_0^{+\infty}\int_{\bbR}\exp\left\{-\frac{\hat c p^2
    t}{2}\right\}\phi(t)\bar J_1(p)\bar J_2(p)dt dp,
     \end{split}
\end{equation}
for any $\phi\in L^1[0,+\infty)$, $J_1,J_2\in{\cal S}$.
Here $\hat c$ is as in Theorem \ref{thm-pinned}.
 \end{thm}

\subsubsection{Case of a no pinning potential}

The result in this case can be formulated as follows.
\begin{thm}
 \label{superdiffusive-1a}
 If   $\hat \al(0)=0$ and $\delta=(3-s)/2$, where $s\in[0,1)$, then
 \begin{equation}
\label{wigner-evolve-2}
\begin{split}
  \lim_{\eps\to0+}& \int_0^{+\infty}\phi(t)C_\eps(t;J_1,J_2)dt\\
  &= {\cal E}_0^2 \int_0^{+\infty}\int_{\bbR}\exp\left\{-\hat c |p|^{3/2}
    t\right\}\phi(t)\bar J_1(p)\bar J_2(p)dt dp,
\end{split}
\end{equation}
for any $\phi\in L^1[0,+\infty)$, $J_1,J_2\in{\cal S}$.  Here $\hat c$
is given by \eqref{hatc}. 
 \end{thm}
The proofs Theorems \ref{diffusive-1a} and  \ref{superdiffusive-1a} are presented in Section \ref{diff-scale-1}.

\subsubsection{Energy fluctuations}

Applying the Wigner fluctuating field to a function $J(y)$
constant in $k$ we obtain the fluctuation field  
\begin{equation}
\label{wigner-fluct-1}
\widetilde W_\eps(t;J) = \sqrt \eps \sum_x \left(\frac
  12|\psi_x^{(\eps)}(t)|^2 - \cal E_0\right) J(\eps x) .
\end{equation}
Denote the empirical fluctuation of energy field  by
\begin{equation}
\label{energy-fluct-1}
{\cal E}_\eps(t;J):=\sqrt \eps \sum_x \left({\frak e}_x
  \left(\frac{t}{\eps^{\delta}}\right) - \cal E_0\right) J(\eps x),\quad
\,J\in C_0^\infty(\bbR) 
\end{equation}
and the respective second mixed moment by
 \begin{equation}
\label{energy-cov}
C_\eps^{(e)}(t;J_1,J_2):= \bbE\left[{\cal E}_\eps(t;J_1){\cal E}_\eps(0;J_2)\right],\quad \,J_1,J_2\in C_0^\infty(\bbR).
\end{equation}
Our next result shows the  fields defined by \eqref{wigner-fluct-1}
and \eqref{energy-fluct-1}  are asymptotically equal. Its proof is presented in Section \ref{sec13.3}.
\begin{prop}
\label{prop012204}
  For any $t\ge0$ we have
\begin{equation}
\label{011704}
\lim_{\eps\to0+}\bbE\left[{\cal E}_\eps(t,J)-
\widetilde{\cal W}_\eps(t;J)\right]^2=0, \quad \,J\in C_0^\infty(\bbR).
\end{equation}
\end{prop}

%
%
As a result the conclusions of Theorems \ref{diffusive-1a} and \ref{superdiffusive-1a} hold for 
$C_\eps^{(e)}(t;J_1,J_2)$ substituted in place of
$C_\eps(t;J_1,J_2)$, which in turn implies  Theorem \ref{thm-fluct-heat}.


\section{Outline of the proofs of Theorems
\ref{diffusive-1} and \ref{superdiffusive-1}}

This section is intended to outline the proof of Theorem
\ref{superdiffusive-1} (Theorem \ref{diffusive-1} follows from a similar consideration). First, in Section
\ref{sec5}, we describe 
the evolution of the Wigner transform $W_\eps(t,y,k)$ of the wave function $\psi^{(\eps)}_x(t)$ introduced in
Section \ref{sec5.0}.  In fact, for our purposes  it is more convenient 
to deal with its Fourier transform in the spatial domain, 
given by \eqref{011507}. 
It satisfies the following equation
 \begin{eqnarray}
\label{out01}
&&\partial_t\widehat W_\eps(t,p,k)=\left(-\frac{i\om'(k)p}{\eps^{\delta-1}}
+\frac{\ga}{\eps^{\delta}}{\cal
  L}\right)\widehat W_\eps(t,p,k) 
\nonumber\\
&&
+\frac{i\ga R'(k)p}{\eps^{\delta-1}}\widehat U_{\eps,-}(t,p,k)-\frac{\ga}{\eps^{\delta}}{\cal L}
\widehat U_{\eps,+}(t,p,k)+O(\eps)
\end{eqnarray}
Here, $\ga=\ga_0\eps^s$ for some $s\in[0,1)$, with  $\delta=(3-s)/2$ and $O(\eps)$ is some expression that becomes negligible, as $\eps\to0+$.  
Here $\widehat U_{\eps,+}(t,p,k)$ represents the  difference between
the kinetic and potential energy,
while $\widehat U_{\eps,-}(t,p,k)$ is related to the energy current
(the product of the momentum and inter-particticle distance). They are
highly oscillatory and  their averages (in time and in $k$)
turn out to vanish in the limit as $\eps\to0+$.

 To  simplify the presentation we assume also here
 that the scattering kernel equals  $R(k,k')=R(k)R(k')$, where $R(k) =
 2 \sin^2(\pi k)$, which is actually
the case for the random momentum exchange model
described in Section \ref{sec-ex}. The scattering operator (see \eqref{scat}) is then of  the form
\begin{equation}
\label{scat-simp}
{\cal L}f(k)=2R(k)\langle f,R\rangle_{L^2(\bbT)}-2 R(k)f(k).
\end{equation}

Since the  wave function at time $t=0$ is $L^2$ bounded, see \eqref{finite-energy}, this bound persists in time, due to the energy conservation. 
  In turn this implies the bound on the norm of $\left(
W_{\eps}(\cdot)\right)$ in $L^{\infty}([0,+\infty);{\cal A}')$, see
\eqref{W-Y}. In consequence  this family  is compact in the 
$*$-weak topology in $L^{\infty}([0,T];{\cal A}')$ for any $T>0$. Our goal is to identify  its
limit as the  function $W(t)$ appearing  in
the statement of Theorem
\ref{superdiffusive-1}. To do so we modify
the argument put forward in \cite{mmm}.   To  further simplify  our presentation we drop the negligible term appearing on the 
right hand side of \eqref{out01}. Performing the Laplace transform on both sides of
\eqref{out01} and using the formula \eqref{scat-simp} for the scattering operator ${\cal
  L}$ we obtain that (dropping the arguments $(\la,p,k)$ to abbreviate the notation)
\begin{eqnarray}
\label{out02}
&&\left(\la +\frac{2\ga}{\eps^{\delta}}R+\frac{i\om'p}{\eps^{\delta-1}}
\right) \bar w_\eps
-\widehat W^{(\eps)}_0 \approx\frac{2\ga R}{\eps^{\delta}}\langle \bar w_\eps,R\rangle_{L^2(\bbT)}
\nonumber\\
&&
+\frac{i\ga R'p}{\eps^{\delta-1}}\bar u_{\eps,-} +\frac{2\ga R}{\eps^{\delta}}
\bar u_{\eps,+} -\frac{2\ga R}{\eps^{\delta}}
\langle\bar u_{\eps,+},R\rangle_{L^2(\bbT)}
\end{eqnarray}
Here $\bar w_{\eps}(\la,p,k)$
and 
$\bar u_{\eps,\pm}(\la,p,k)$ are  the
Laplace transforms of  $\widehat W_{\eps}(t,p,k)$
$\widehat U_{\eps,\pm}(t,p,k)$, respectively, see \eqref{exp-wigner-eqt-1k}.  After some simple computations we get  
\begin{equation}
\label{out03}
\bar w_\eps-\frac{2\ga R}{D^{(\eps)}}\langle  \bar w_\eps,R\rangle_{L^2(\bbT)}
-\frac{\eps^{\delta}\widehat W^{(\eps)}_0}{D^{(\eps)}}\approx
\frac{i\ga \eps R'p}{D^{(\eps)}}\bar u_{\eps,-}+\frac{2\ga R}{D^{(\eps)}}
\bar u_{\eps,+}
-\frac{2\ga R}{D^{(\eps)}}\langle
\bar u_{\eps,+},R\rangle_{L^2(\bbT)} ,\end{equation}
where $D^{(\eps)}:=\la\eps^{\delta}+2\ga R+i\eps \om'$.
Performing the scalar product of both sides of the equation against $2\ga R/\eps^{\delta}$ we conclude 
\begin{equation}
\label{out04}
a_w^{(\eps)}\langle  \bar w_\eps,R\rangle_{L^2(\bbT)}
-\int_{\bbT}\frac{2\ga R\widehat W^{(\eps)}_0}{D^{(\eps)}}dk\approx
{\cal O}_\eps ,\end{equation}
where $\widehat W^{(\eps)}_0$ is the Fourier-Wigner transform of the initial condition and
$$
a_w^{(\eps)}:=\frac{2\ga}{\eps^{\delta}}\left(1-\int_{\bbT}\frac{2\ga R^2}{D^{(\eps)}}dk\right).
$$
Here ${\cal O}_\eps$ is the expression that arises from the scalar multiplication of the right hand side of \eqref{out03}.

It is quite simple to show that the second term on the left hand side of \eqref{out04} tends to $\overline W_0(p)$, given by \eqref{W0}.
Our main effort goes into proving that the right hand side of \eqref{out04} vanishes as $\eps\to0+$
and that 
$$
a_w^{(\eps)}\to \la +C |p|^{3/2}
$$ 
for an appropriate $C>0$, as $\eps\to0+$, when $\om(0)=0$ (we have $a_w^{(\eps)}\to \la +C p^{2}$ in the unpinned case). The first fact is a consequence of the aforementioned oscillatory behavior of $\bar u_{\eps,\pm}$, while the convergence of $a_w^{(\eps)}$
follows from detailed calculations, see  Proposition \ref{cor012301}. This allows us to conclude that  $\langle   w(\la,p),R\rangle_{L^2(\bbT)}$ - the  limit of  $\langle   w_\eps(\la,p),R\rangle_{L^2(\bbT)}$, as $\eps\to0+$, satisfies 
\begin{equation}
\label{011302-15}
( \la +C |p|^{3/2})\langle   w(\la,p),R\rangle_{L^2(\bbT)}=\overline{ W}_0(p). 
\end{equation}
Since in the macroscopic time the number of random collisions grows as $\eps^{s-\delta}\gg1$ (recall that is proportional to $\ga/\eps^{\delta}\sim\eps^{s-\delta}$) the limit  $w(\la,p,k)$  of energy density  $\bar w_{\eps}(\la,p,k)$ for a fixed $p$, as $\eps\to0+$, should become independent of the $k$-variable, therefore, since $\int_{\bbT}R(k)dk=1$, we ought to have
\begin{equation}
\label{031302-15}
 w(\la,p,k)\equiv  w(\la,p)=\langle   w(\la,p),R\rangle_{L^2(\bbT)}.
\end{equation}
This {\em homogenization} result is proved in Theorem \ref{cor011811a} and allows us to conclude \eqref{031302-15}. By virtue of \eqref{011302-15} and the uniqueness property of the Laplace transform we infer that $W(t)$, appearing in the statement of Theorem \ref{superdiffusive-1}, satisfies \eqref{superheat1c}.
The  ''true'' argument is a bit more involved, due to the fact that the scattering kernel
$R(k,k')$ corresponding to  the noise considered in this paper is not of a product type, see \eqref{060411}, which complicates the actual calculations.

\bigskip 
\section{Evolution of the  Wigner transform}

\label{sec5}

 For a  given $\eps>0$ let $\hat \psi^{(\eps)}(t)$ be a solution of  \eqref{basic:sde:2} with the initial condition  $\hat \psi$ distributed according to a probability measure $\mu_\eps$ on $L^2(\bbT)$.
The Fourier transform of the Wigner transform of  $\hat \psi^{(\eps)}(t)$ is given by
\begin{equation}
\label{011507}
\widehat W_{\eps}(t,p,k):=\frac{\eps}{2}\bbE_\eps\left[\left(\hat \psi^{(\eps)}\right)^*\left(t,k-\frac{\eps p}{2}\right)\hat \psi^{(\eps)}\left(t,k+\frac{\eps p}{2}\right)\right],
\end{equation}
where,  as we recall, $\bbE_{\eps}$ is the average with respect to the initial
condition and the realization of the noise.
To close the equations governing the dynamics of $\widehat
W_{\eps}(t,p,k)$ we shall also need the following functions
 \begin{eqnarray}
 \label{021507}
&&
\widehat Y_{\eps}(t,p,k):=\frac{\eps}{2}\bbE_\eps\left[\hat \psi^{(\eps)}\left(t,- k+\frac{\eps p}{2}\right)\hat \psi^{(\eps)}\left(t, k+\frac{\eps p}{2}\right)\right], \\
&&\widehat Y_{\eps,-}(t,p,k) :=\widehat Y_\eps^*(t,-p,k),\quad
\widehat W_{\eps,-}(t,p,k) :=\widehat W_\eps(t,p,-k). \nonumber
\end{eqnarray}
We shall also write  $\widehat W_{\eps,+}=\widehat W_{\eps}$ and
$\widehat Y_{\eps,+}=\widehat Y_{\eps}$.
Define  (cf \eqref{r})
\begin{equation}
\label{R}
R(k,k',p):=\frac12\sum_{\si=\pm1}r\left(k-\frac{ p}{2},k-\si k'\right)r\left(k+\frac{ p}{2},k-\si k'\right).
\end{equation}
With the above definition we can write (see \eqref{R-k} for definition $R(\cdot)$)
 \begin{eqnarray}
\label{exp-wigner-eqt}
&&\partial_t\widehat W_\eps(t,p,k)=-\left[\frac{i}{\eps^{\delta-1}}\delta_{\eps}\om(k;p)+\frac{2\ga\bar R(k,\eps p)}{\eps^{\delta}}\right]\widehat W_\eps(t,p,k)\nonumber\\
&&+\frac{\ga}{\eps^{\delta}}\left\{R\left(k-\frac{\eps p}{2}\right)\widehat  Y_\eps(t, p,k)+R\left(k+\frac{\eps p}{2}\right)\widehat Y_{\eps,-}(t, p,k)\right\}\nonumber \\
&&\!\!\!\!\!\!\!\!\!\!\!\!\!\!\!\\
&&
+\frac{4\ga}{\eps^{\delta}}\int_{\bbT} 
r\left(k-\frac{\eps p}{2},k'\right)
r\left(k+\frac{\eps p}{2},k'\right)\times\nonumber\\ 
&&\qquad\qquad \times\bbE\left[(\hat{\frak p}^{(\eps)})^*\left(t,k-k'-\frac{\eps p}{2}\right)\hat{\frak p}^{(\eps)}\left(t,k-k'+\frac{\eps p}{2}\right)\right]dk',\nonumber
\end{eqnarray}
where  $\delta_{\eps}\om(k,p)$ and $\bar R(k,p)$ are defined in \eqref{010906} 
and
$$
\hat{\frak p}^{(\eps)}(t,k):=\frac{1}{2i}\left[\hat\psi^{(\eps)}(t,k)-(\hat\psi^{(\eps)})^*(t,-k)\right]
$$
is the Fourier transform of the momentum.
Since the latter is  real valued, its Fourier transform is  complex even.
 The last term appearing on the right hand side of
 \eqref{exp-wigner-eqt} can be replaced by  
\begin{equation}
\label{exp-wigner-eqt-10}
\begin{split}
  \frac{\ga}{\eps^{\delta}}\int_{\bbT}R(k,k',\eps p)
  \bbE_\eps\left[(\hat{\psi}^{(\eps)})^*\left(t,k'-\frac{\eps
        p}{2}\right) \hat{\psi}^{(\eps)}\left(t,k'+\frac{\eps
        p}{2}\right)\right]dk'\\
  = \frac{2\ga}{\eps^{\delta}}({\cal R}_{\eps p} \widehat W_\eps) (t,p,k)
\end{split}
\end{equation}
where
\begin{equation}
 \label{010511}
 {\cal R}_p f(k):=\int_{\bbT} R(k,k',p)f(k')dk'.
\end{equation}
Therefore, 
\begin{eqnarray}
\label{exp-wigner-eqt-1}
&&\partial_t\widehat W_\eps(t,p,k)=-\frac{i}{\eps^{\delta-1}}\delta_{\eps}\om(k;p)
\widehat W_\eps(t,p,k)
+\frac{\ga}{\eps^{\delta}}{\cal L}_{\eps p}\widehat W_\eps(t,p,k)\nonumber\\
&&\\
&&
-\frac{\ga}{2\eps^{\delta}}\sum_{\si\in\{-,+\}}{\cal L}^{+}_{\si\eps p}\widehat Y_{\eps,-\si}(t,p,k),\nonumber 
\end{eqnarray}
and
\begin{eqnarray}
\label{050411}
&&
{\cal L}_p f(k):=2{\cal R}_{ p} f(k) - 2\bar R (k,p) f(k),\nonumber\\
&&
\\
&&
{\cal L}_{p}^\pm f(k):=2{\cal R}_{p} f(k) -2R\left(k\pm\frac p2\right)
f(k).\nonumber 
\end{eqnarray}
In addition, 
\begin{eqnarray}
\label{anti-wigner-eqt}
&&
\partial_t\widehat Y_\eps(t,p,k)=-\frac{2i}{\eps^{\delta}} \bar\om(k,\eps p)\widehat Y_\eps(t,p,k)-\frac{2\ga}{\eps^{\delta}} \bar R(k,\eps p)\widehat Y_\eps(t,p,k)
\nonumber\\
&&\\
&&
+\frac{\ga}{\eps^{\delta}}R\left(k-\frac{\eps p}{2}\right)\widehat W_{\eps}(t,p,k)+\frac{\ga}{\eps^{\delta}}R\left(k+\frac{\eps p}{2}\right)\widehat W_{\eps,-}(t,p,k)+\frac{\ga}{\eps^{\delta}}{\cal U}_\eps(t,p,k)\nonumber
\end{eqnarray}
where 
\begin{eqnarray*}
&&{\cal U}_\eps(t,p,k):=
4\int_{\bbT}r\left(-k+\frac{\eps p}{2},k'\right)r\left(k+\frac{\eps p}{2},-k'\right)
\\
&&
\times
\bbE_\eps\left[{\frak p}^{(\eps)}\left(t,-k-k'+\frac{\eps p}{2}\right){\frak p}^{(\eps)}\left(t,k+k'+\frac{\eps p}{2}\right)\right]dk.
\end{eqnarray*}
After straightforward calculations (cf  \eqref{exp-wigner-eqt} -- \eqref{exp-wigner-eqt-1})
we conclude that
\begin{eqnarray*}
&&{\cal U}_\eps(t,p,k)
={\cal R}_{\eps p} \left(\widehat Y_\eps + \widehat Y_{\eps,-}-
  \widehat W_\eps - \widehat W_{\eps,-} \right) (t,p,k).
\end{eqnarray*}
Then,
\begin{eqnarray}
\label{anti-wigner-eqt1}
&&
\partial_t\widehat Y_\eps(t,p,k)=-\frac{2i}{\eps^{\delta}} \bar\om(k,\eps p)\widehat Y_\eps(t,p,k)+\frac{\ga}{\eps^{\delta}}  {\cal L}_{\eps p}\widehat Y_\eps(t,p,k)
\nonumber\\
&&
\\
&&
+\frac{\ga}{\eps^{\delta}} {\cal R}_{\eps p}(\widehat
Y_{\eps,-}-\widehat Y_{\eps} )(t,p,k) 
-\frac{\ga}{2\eps^{\delta}}\sum_{\si\in\{-,+\}} {\cal L}^+_{\si\eps p}
\widehat W_{\eps,-\si}(t,p,k) . \nonumber 
\end{eqnarray}

From \eqref{exp-wigner-eqt-1} and \eqref{anti-wigner-eqt1}
we conclude that for any fixed $p\in\bbR$ the evolution $(\widehat W_\eps(t), \widehat Y_\eps(t),
\widehat Y_{\eps,-}(t), \widehat W_{\eps,-}(t) )$ is governed
by a closed system of four linear equations with a generator that is a bounded operator in
$(L^r(\bbT))^4$ for any $r\in[1,+\infty]$. In particular, under
the assumption that the initial distribution of the wave functions
satisfies \eqref{psi} the components of $(\widehat W_\eps(t), \widehat Y_\eps(t),
\widehat Y_{\eps,-}(t), \widehat W_{\eps,-}(t) )$ belong to $C([0,+\infty);{\cal A}')$.




 \section{Expansion of the dynamics of the Wigner transform}

\label{sec9a}

Assumption \eqref{psi} 
guarantees that 
\begin{equation}
\label{012003}
\sum_{\iota\in\{-,+\}}\sup_{p\in\bbR}(\|\widehat W_{\eps,\iota}(0,p,\cdot)\|_{L^2(\bbT)}+\|\widehat Y_{\eps,\iota}(0,p,\cdot)\|_{L^2(\bbT)})<+\infty.
\end{equation}
From an elementary existence and uniqueness result concerning the dynamics 
of $(\widehat W_\eps(t), \widehat Y_\eps(t),
\widehat Y_{\eps,-}(t), \widehat W_{\eps,-}(t) )$ in
$(L^2(\bbT))^4$ we conclude that  
\begin{equation}
\label{012003a}
\sup_{t\in[0,T]}\sum_{\iota\in\{-,+\}}(\|\widehat W_{\eps,\iota}(t,p,\cdot)\|_{L^2(\bbT)}+\|\widehat Y_{\eps,\iota}(t,p,\cdot)\|_{L^2(\bbT)})<+\infty.
\end{equation}
for any $\eps,T>0$ and $p\in\bbR$. 

 We expand the scattering kernel appearing on the
 right hand side of  \eqref{exp-wigner-eqt-1},
 \eqref{anti-wigner-eqt1} into the powers of $\eps$, up to the second order. 
 To abbreviate the notation we shall write 
\begin{equation}
\label{022312}
 R_\eps:=R(k)+\frac{(\eps
      p)^2}{8}R''(k),\quad \bar\om:=\bar\om(k,\eps p),\quad
    \delta_\eps\om:=\delta_\eps\om(k, p) .
\end{equation}
Since $\partial_p R(k, k', 0)= 0$ we can write
 \begin{eqnarray}
\label{exp-wigner-eqt-1c}
&&\partial_t\widehat W_\eps=-\frac{i\delta_\eps\om}{\eps^{\delta-1}}
\widehat W_\eps+
\left(\frac{\ga}{\eps^{\delta}}{\cal L}+\frac{\ga p^2}{2\eps^{\delta-2}}(\delta^2{\cal L})\right)\left[\widehat W_\eps-\frac{1}{2}\left(\widehat Y_\eps+
\widehat Y_{\eps,-}\right)\right]\nonumber\\
&&\\
&&
+\frac{\ga R'p}{2\eps^{\delta-1}}(\widehat Y_{\eps,-}-\widehat Y_{\eps})
+\eps^{3-\delta}{\frak R}^{(1)}_\eps,\nonumber\nonumber
\end{eqnarray}
and
 \begin{eqnarray}
 \label{anti-wigner-eqt1d}
&&
\partial_t\widehat Y_\eps=-\frac{2i\bar\om}{\eps^{\delta}}
\widehat Y_\eps+\left(\frac{\ga}{\eps^{\delta}}
{\cal L}+
 \frac{\ga p^2}{2\eps^{\delta-2}}( \delta^2 {\cal L})\right)\left[\widehat Y_\eps
-\frac{1}{2}\left(\widehat W_\eps+\widehat W_{\eps,-}\right)\right]
\nonumber\\
&&
\\
&&
+\left(\frac{\ga}{\eps^{\delta}} {\cal R}+\frac{\ga p^2}{2\eps^{\delta-2}} (\delta^2{\cal R})\right)(\widehat Y_{\eps,-}-\widehat Y_{\eps})
+\frac{\ga R'p}{2\eps^{\delta-1}}(\widehat
W_{\eps,-}-\widehat W_{\eps})
 +\eps^{3-\delta}{\frak R}^{(2)}_\eps,
\nonumber
\end{eqnarray}
where ${\cal L}$, $ {\cal R}f$ are given by \eqref{050411} and  \eqref{010511} and 
 \begin{eqnarray}
\label{L-2}
&&
(\delta^2{\cal L})f(k):=-\pi^2\int_{\bbT}R_1(k,k') f(k') dk'-\frac{1}{2}R'' (k) f(k),\\
 &&
 (\delta^2 {\cal R})f(k):=\int_{\bbT} \partial_p^2R(k,k',0)f(k')dk'=-\frac{\pi^2}{2}\int_{\bbT}R_1(k,k') f(k') dk'.\nonumber
\end{eqnarray}
Here
\begin{equation}
\label{032110a}
R_1(k,k'):=-\frac{2}{\pi^2}\partial_p^2 R(k,k',0)
=4 \frak f_+(k) \frak f_{+}(k')+\frak f_+(k) \frak e_{-}(k')+3\frak f_{-}(k)  \frak e_{+}(k'),
\end{equation}
with $\frak{e}_\pm(k)$ defined in \eqref{frak-e}
and
\begin{equation}
\label{011701}
\frak{f}_+(k):=2{\frak s}^2( k),\quad \frak{f}_{-}(k):=2{\frak c}^2( k).
\end{equation}
 Note that
\begin{equation}
\label{020912}
(\delta^2{\cal L})f_-(k)=\left((\delta^2{\cal L})f\right)(-k),
\end{equation}
where $f_-(k):=f(-k)$.
In addition, 
\begin{equation}
 \label{null}
{\frak R}^{(i)}_\eps(t,0,k)\equiv0,\quad i=1,2.
 \end{equation}
Adopting the convention $\widehat W_{\eps}=\widehat W_{\eps,+}$,  $\widehat Y_{\eps}=\widehat Y_{\eps,+}$,  we can write that  for any $M>0$ 
\begin{equation}
 \label{F}
 \|{\frak R}^{(i)}_\eps(t,p,\cdot)\|_{L^2(\bbT)}\preceq \sum_{\iota\in\{-,+\}}(\|\widehat W_{\eps,\iota}(t,p,\cdot)\|_{L^2(\bbT)}+\|\widehat Y_{\eps,\iota}(t,p,\cdot)\|_{L^2(\bbT)})
 \end{equation}
 for $i=1,2$ and  all $t\ge0$, $\eps\in(0,1]$, $|p|\le M$.
Define 
\begin{eqnarray}
\label{U-V}
&&
\widehat U_{\eps,+}(t,p,k):= \frac{1}{2}(\widehat Y_{\eps}+\widehat Y_{\eps,-}) (t,p,k),\\
&&
\widehat U_{\eps,-}(t,p,k):= \frac{1}{2i}(\widehat Y_{\eps}-\widehat Y_{\eps,-}) (t,p,k).\nonumber
\end{eqnarray}
From \eqref{exp-wigner-eqt-1c}
and \eqref{anti-wigner-eqt1d} we get
 \begin{eqnarray}
\label{exp-wigner-eqt-1ca}
&&\partial_t\widehat W_{\eps,+}=-\frac{i\delta_\eps\om}{\eps^{\delta-1}}
\widehat W_{\eps,+}-\frac{i\ga R'p}{\eps^{\delta-1}}\widehat U_{\eps,-}
+\left(\frac{\ga}{\eps^{\delta}}{\cal L}\nonumber 
+\frac{\ga p^2}{2\eps^{\delta-2}}(\delta^2{\cal L})\right)\left(\widehat W_{\eps,+}-U_{\eps,+}\right)
+\eps^{3-\delta}\bar {\frak R}^{(1)}_\eps,\nonumber
\\
&&
\\
&&
\partial_t\widehat U_{\eps,+}=\frac{2\bar\om}{\eps^{\delta}}
\widehat U_{\eps,-}+\left(\frac{\ga}{\eps^{\delta}}
{\cal L}+
 \frac{\ga p^2}{2\eps^{\delta-2}}( \delta^2 {\cal L})\right)\left[\widehat U_{\eps,+}
-\frac{1}{2}\left(\widehat W_{\eps,+}+\widehat W_{\eps,-}\right)\right]
+\eps^{3-\delta}\bar{\frak  R}^{(2)}_\eps,
\nonumber
\\
&&
\nonumber\\
&&
\partial_t\widehat U_{\eps,-}=-\frac{2\bar\om}{\eps^{\delta}}
\widehat U_{\eps,+}
-\frac{2\ga}{\eps^{\delta}}R_\eps\widehat U_{\eps,-}
-\frac{i\ga R'p}{2\eps^{\delta-1}}(\widehat
W_{\eps,-}-\widehat W_{\eps,+})
 +\eps^{3-\delta}\bar{\frak R}^{(3)}_\eps,
\nonumber\\
&&
\nonumber\\
&&\partial_t\widehat W_{\eps,-}=\frac{i\delta_\eps\om}{\eps^{\delta-1}}
\widehat W_{\eps,-}+\frac{i\ga R'p}{\eps^{\delta-1}}\widehat U_{\eps,-}
+\left(\frac{\ga}{\eps^{\delta}}{\cal L}\
+\frac{\ga p^2}{2\eps^{\delta-2}}(\delta^2{\cal
  L})\right)\left(\widehat W_{\eps,-}-\widehat U_{\eps,+}\right)
+\eps^{3-\delta}\bar {\frak R}^{(4)}_\eps,\nonumber
\end{eqnarray}
From \eqref{F} we conclude that for any $M>0$, $i\in\{1,2,3,4\}$
 \begin{equation}
 \label{F1}
 \|\bar{\frak R}^{(i)}_\eps(t,p,\cdot)\|_{L^2(\bbT)}\preceq \sum_{\iota\in\{-,+\}}(\|\widehat W_{\eps,\iota}(t,p,\cdot)\|_{L^2(\bbT)}+\|\widehat U_{\eps,\iota}(t,p,\cdot)\|_{L^2(\bbT)})
 \end{equation}
 for $t\ge0$ and $\eps\in(0,1]$, $|p|\le M$.

Let ${\cal D}(\phi):={\cal D}(\phi,\phi)$, where
\begin{equation*}
\begin{split}
&{\cal D}(\phi,\psi):=\int_{\bbT}(-{\cal L})\phi(k)\psi^*(k)dk\\
&=\int_{\bbT^2}R(k,k')[\phi(k)-\phi(k')][\psi(k)-\psi(k')]^*dkdk',
\end{split}
\end{equation*}
for any $ \phi,\psi\in L^2(\bbT)$ and
$$
{\frak E}_\eps(t,p):=\sum_{\iota\in\{-,+\}}\left(\frac12\|\widehat W_{\eps,\iota}(t,p)\|_{L^2(\bbT)}^2+\|\widehat U_{\eps,\iota}(t,p)\|_{L^2(\bbT)}^2\right).
$$
Taking the scalar products of both sides of equations appearing in
\eqref{exp-wigner-eqt-1ca} against the respective $\widehat
W_{\eps,\iota}$, $\widehat U_{\eps,\iota}$, $\iota\in\{-,+\}$ we
obtain 
\begin{eqnarray}
\label{011501}
&&
\frac12{\frak E}_\eps(t,p)+\frac{\ga}{\eps^{\delta}}\int_0^t{\cal
  D}\left((\widehat W_{\eps}-\widehat
  U_{\eps,+})(s,p)\right)ds\nonumber\\
&&+\frac{2\ga}{\eps^{\delta}}\int_0^tds\int_{\bbT}R(k)|\widehat U_{\eps,-}(s,p,k)|^2dk\\
&&
+ 2\eps^{1-\delta}\ga p\int_0^tds\int_{\bbT} R'(k){\rm
  Im}\left(\widehat U_{\eps,-}^*\widehat
  W_{\eps,+}\right)(s,p,k)dk\nonumber\\
&&
 \quad\quad = \frac12{\frak E}_\eps(0,p)
+\eps^{2-\delta}\int_0^t{\frak R}_\eps(s,p)ds,\nonumber 
\end{eqnarray}
where for any $M>0$ we have
 \begin{equation}
 \label{F2}
 {\frak R}_\eps(t,p)\preceq {\frak E}_\eps(t,p), \quad t\ge0, \,|p|\le M,\,\eps\in(0,1].
 \end{equation}
Using Young's inequality and the
 fact that $ (R'(k))^2\preceq R(k)$ (see \eqref{beta1} below) we
 conclude that for any $M>0$ there exists
$C>0$ such that
 \begin{eqnarray}
 \label{011601}
&&
\eps^{1-\delta}\ga p\int_0^tds\int_{\bbT}R'(k){\rm Im}\left(\widehat U_{\eps,-}^*\widehat W_{\eps,+}\right)(s,p,k)dk\\
&&
\ge
-\frac{\ga}{\eps^{\delta}}\int_0^tds\int_{\bbT}R(k)|\widehat U_{\eps,-}(s,p,k)|^2dk
-C\eps^{2-\delta}\int_0^t\|\widehat W_{\eps,+}(s,p)\|_{L^2(\bbT)}^2ds,\nonumber
\end{eqnarray}
for  $t\ge0$, $|p|\le M$ and $\eps\in(0,1]$.
From the above, estimate \eqref{F1}, identity \eqref{011501} and Gronwall's inequality we obtain the following.
\begin{prop}
\label{prop1}
For any $M>0$ there exists $C_1>0$ such that
 \begin{equation}
 \label{021501}
 {\frak E}_\eps(t,p)\le {\frak E}_\eps(0,p)e^{C_1\eps^{2-\delta}
   t},\quad\forall\,\eps\in(0,1],\,|p|\le M,\,t\ge0.
 \end{equation}
\end{prop}


\section{Laplace transform of  system (\ref{exp-wigner-eqt-1ca})}
 \label{sec-homo}

 For any $\la> \la_0^{(\eps)}:=C_1\eps^{2-\delta}$ ($C_1$ as in
 Proposition \ref{prop1}) we let
\begin{eqnarray}
\label{lafo}
&&
\bar w_{\eps,\iota}(\la,p,k):=\int_0^{+\infty} e^{-\la t}\widehat W_{\eps,\iota}(t,p,k)dt,\\
&&
\nonumber\\
&&
\bar u_{\eps,\iota}(\la,p,k):=\int_0^{+\infty} e^{-\la t}\widehat U_{\eps,\iota}(t,p,k)dt.\nonumber
\end{eqnarray}
Thanks to Proposition \ref{prop1} the above integrals are well defined  in
$L^2(\bbT)$ for  $\la> \la_0^{(\eps)}$ and $p\in\bbR$. 
 Let
\begin{equation}
\label{l-0}
\la_0:=\left\{
\begin{array}{ll}
0,&\delta<2\\
C_1,& \delta=2,
\end{array}
\right.
\end{equation} 
and $C_1$ is as in \eqref{021501}. 
The ``remainder'' term $\bar r^{(i)}_\eps(\la,p,k)$, that is the Laplace transforms of 
$\bar{\frak R}^{(i)}_\eps(t,p,k)$, has the following property: for any $M>0$ and compact interval $I\subset (\la_0,+\infty)$  
\begin{equation}
 \label{LF1}
 \|\bar r^{(i)}_\eps(\la,p)\|_{L^2(\bbT)}\preceq \sum_{\iota\in\{-,+\}}(\|\bar w_{\eps,\iota}(\la,p)\|_{L^2(\bbT)}+\|\bar u_{\eps,\iota}(\la,p)\|_{L^2(\bbT)})
 \end{equation}
 for $i=1,2,3,4$ and $\eps\in(0,1]$, $|p|\le M$, $\la\in I$.
Therefore,
 from Proposition \ref{prop1} we conclude that 
  \begin{equation}
 \label{C-I}
 C_I:=\sup_{\eps\in(0,1]}\sup_{\la\in I,|p|\le M}\left(\|\bar w_{\eps,\iota}(\la,p)\|_{L^2(\bbT)}+\sum_{\iota\in\{-,+\}}\|\bar u_{\eps,\iota}(\la,p)\|_{L^2(\bbT)}\right)<+\infty.
 \end{equation}
Taking the Laplace transform of the both sides of
equations of the system \eqref{exp-wigner-eqt-1ca}
we obtain
 \begin{eqnarray}
\label{exp-wigner-eqt-1k}
&&\la \bar w_{\eps,+}-\widehat W_{\eps,+}^{(0)}=-\frac{i\delta_\eps\om}{\eps^{\delta-1}}
\bar w_{\eps,+}-\frac{i\ga R'p}{\eps^{\delta-1}}\bar u_{\eps,-}
 +\frac{\ga}{\eps^{\delta}}L_\eps
\left(\bar w _{\eps,+}-\bar u_{\eps,+}\right)
+\eps^{3-\delta}\bar r^{(1)}_\eps,\nonumber
\\
&&
\\
&&
\la \bar u_{\eps,+}-\widehat U_{\eps,+}^{(0)}=\frac{2\bar\om}{\eps^{\delta}}
\bar u_{\eps,-}
+\frac{\ga}{\eps^{\delta}} L_\eps\left[\bar u_{\eps,+}-\frac{1}{2}\left(\bar w_{\eps,+}+\bar w_{\eps,-}\right)\right]
+\eps^{3-\delta}\bar r^{(2)}_\eps,
\nonumber
\\
&&
\nonumber\\
&&
\la \bar u_{\eps,-}-\widehat U_{\eps,-}^{(0)}=-\frac{2\bar\om}{\eps^{\delta}}
\bar u_{\eps,+}
-\frac{2\ga}{\eps^{\delta}}R_\eps\bar u_{\eps,-}
-\frac{i\ga R'p}{2\eps^{\delta-1}}(\bar
w_{\eps,-}-\bar w_{\eps,+})
 +\eps^{3-\delta}\bar r^{(3)}_\eps
\nonumber\\
&&
\nonumber\\
&&\la \bar w_{\eps,-}-\widehat W_{\eps,-}^{(0)}=\frac{i\delta_\eps\om}{\eps^{\delta-1}}
\bar w_{\eps,-}+\frac{i\ga R'p}{\eps^{\delta-1}}\bar u_{\eps,-}
+\frac{\ga}{\eps^{\delta}}L_\eps\left(\bar w_{\eps,-}-\bar u_{\eps,+}\right)
+\eps^{3-\delta}\bar r^{(4)}_\eps,\nonumber
\end{eqnarray}
where
$
L_\eps:={\cal L}
+(1/2) (\eps p)^2(\delta^2{\cal L})
$
and ${\cal L}$, $\delta^2{\cal L}$ are given by \eqref{scat} and  \eqref{L-2}, respectively.
Performing the real parts of the scalar products  in $L^2(\bbT)$ of  the respective  equations of the above system with   $(1/2)\bar w_{\eps,\pm}(\la,p,k)$, $ \bar u_{\eps,\pm}(\la,p,k)$ and adding them sideways we get
\begin{eqnarray}
\label{exp-wigner-eqt-1z}
&&\la\left(\| \bar w_{\eps}(\la,p)\|_{L^2(\bbT)}^2+\sum_{\iota\in\{-,+\}}\| \bar u_{\eps,+}(\la,p)\|_{L^2(\bbT)}^2\right)\nonumber\\
&&
+\frac{2\ga}{\eps^{\delta}}
\int_{\bbT}R_\eps(k)|\bar u_{\eps,-}(\la,p,k)|^2dk+2\eps^{1-\delta}\ga
p \int_{\bbT} R'(k){\rm Im} \left(\bar u_{\eps,-}^*
\bar w_{\eps}\right)(\la,p,k)\; dk\nonumber\\ 
&&
+\frac{\ga}{\eps^{\delta}}{\cal D}\left((\bar w_{\eps}-\bar
  u_{\eps,+})(\la,p)\right)={\rm Re} \int_{\bbT} \widehat
W_{\eps}^{(0)}(p,k)\bar w_{\eps}(\la,p,k)\; dk
\\
&&
+{\sum_{\iota\in\{-,+\}}\rm Re} \int_{\bbT} \widehat
U_{\eps,\iota}^{(0)}(p,k) \bar u_{\eps,\iota}(\la,p,k) \; dk 
+\bar {\frak R}_\eps(\la,p).
\nonumber
\end{eqnarray}
Given $M>0$ and $I\subset (\la_0,+\infty)$ compact, we have
$$
\bar {\frak R}_\eps(\la,p)\preceq \|\bar w_{\eps,\iota}(\la,p)\|_{L^2(\bbT)}^2+\sum_{\iota\in\{-,+\}}\|\bar u_{\eps,\iota}(\la,p)\|_{L^2(\bbT)}^2
$$
for all $\eps\in(0,1]$ and $\la\in I$, $|p|\le M$. Using again Young's inequality, as in \eqref{011601},
together with \eqref{C-I} we conclude that for any $M>0$, a compact
interval $I\subset (\la_0,+\infty)$ and  $\eps\in(0,1]$
\begin{equation}
\label{021601c}
\sup_{\la\in I,|p|\le M}\left[\int_{\bbT}R_\eps(k)|\bar u_{\eps,-}(\la,p,k)|^2dk+{\cal D}\left((\bar w_{\eps}-\bar u_{\eps,+})(\la,p)\right)\right]\preceq \eps^{\delta-s}.
\end{equation}
{In fact it is possible to get a more precise result.
\begin{prop}
\label{prop011601}
 For any $M>0$ and a compact interval $I\subset (\la_0,+\infty)$
\begin{eqnarray}
\label{021601}
&&\sup_{\la\in I,|p|\le M}\left[\int_{\bbT}R_\eps(k)|\bar
  u_{\eps,-}(\la,p,k)|^2dk+ 
{\cal D}\left(\bar w_{\eps}(\la,p)\right)\right.\\
&&
\left.+\int_{\bbT} R_\eps(k)|\bar u_{\eps,+}(\la,p,k)|^2 
\left(\frac{ \bar\om(k,\eps p)}{\la \eps^{\delta}+\ga R_\eps(k)}\right)^2dk\right]
\preceq \eps^{\delta-s},\quad \eps\in(0,1].\nonumber
\end{eqnarray}
\end{prop}
We postpone the proof of the above Proposition  till Section \ref{sec14.2} and use
it first to show a homogenization result formulated below.
Define  
\begin{equation}
\label{010301-15}
 w_{\eps}^{(\iota)}(\la,p) :=\int_{\bbT}\bar  w_{\eps}(\la,p,k){\frak e}_{\iota}(k)
 dk ,\quad \iota\in\{-,+\}.
 \end{equation}
\begin{thm}
\label{cor011811a} 
Suppose that the initial laws $(\mu_\eps)$ satisfy \eqref{psi}. Then,
for any $M>0$ and   a compact interval $I\subset (\la_0,+\infty)$ we have
\begin{equation}
\label{031811c}
\lim_{\eps\to0+}\sup_{\la\in I,|p|\le M}\int_{\bbT} \left|\bar w_{\eps}(\la,p,k)-
  w_{\eps}^{(\pm)}(\la,p)\right| dk=0,
\end{equation}
and
\begin{equation}
\label{031811d}
\lim_{\eps\to0+}\sup_{\la\in I,|p|\le M} \int_{\bbT} \left| \bar
  u_{\eps,\iota}(\la,p,k)\right| dk = 0,\quad \iota\in\{-,+\}.
\end{equation}
\end{thm}
\proof
Consider the case $\iota = -$.  By \eqref{C-I}, it is enough to prove that 
 \begin{equation}
\label{031811cc}
\lim_{\eps\to0+}\sup_{\la\in I,|p|\le M}\int_{\bbT} \left|\bar w_{\eps}(\la,p,k)-
  w_{\eps}^{(-)}(\la,p)\right|{\frak e}_{-}(k) dk=0.
\end{equation}
 By virtue of 
 \eqref{021601} we have
\begin{eqnarray*}
&&
\int_{\bbT}\left| \bar w_{\eps}(\la,p,k)-
  w_{\eps}^{(-)}(\la,p) \right|\frak e_{-}(k)dk\\
&&
\le \int_{\bbT^2}\left| \bar w_{\eps}(\la,p,k)-
 \bar w_{\eps}(\la,p,k')
\right|\frak e_{-}(k)\frak e_{-}(k')dk dk'\\
&&
\le \left\{\int_{\bbT^2}R(k,k')\left| \bar w_{\eps}(\la,p,k)-
\bar  w_{\eps}(\la,p,k')
\right|^2dk dk'\right\}^{1/2}
\\
&&
\qquad \qquad \times\left\{\int_{\bbT^2}\frac{ \frak e^2_{-}(k)\frak e^2_{-}(k')}{R(k,k')}dkdk'\right\}^{1/2}
\preceq \eps^{\delta/2}.
\end{eqnarray*}
The last estimate follows from (see
\eqref{060411})
$$
\frac{ \frak e^2_{-}(k)\frak e^2_{-}(k')}{R(k,k')}\preceq \frac{ \frak
  e_{-}(k)\frak e_{-}(k')}{ \frak e_{-}(k)+\frak e_{-}(k')}\preceq 1, 
$$
 This together with \eqref{C-I} imply
\eqref{031811c}.
The case $\iota = +$ can be argued similarly.
The proof of \eqref{031811d} is a consequence of \eqref{C-I} and
\eqref{021601}.

\qed


\section{Identification of the limit of the Wigner transform}

\label{sec-iden}

Recall that $(W_\eps(\cdot))$ is $*-$weakly
sequentially compact in  $L^\infty([0,T],{\cal A}')$ 
for any $T>0$. Therefore for any $\eps_n\to 0$, as
$n\to+\infty$, we can choose a subsequence, denoted in the same way,
such that it $*-$weakly 
converges to some $W(\cdot) \in L^\infty([0,T],{\cal A}')$.
In light of \eqref{W-Y} we have
\begin{equation}
\label{W-Y1}
\sup_{t\ge0}\| W(t)\|_{{\cal A}'}\le K_0,
\end{equation}
with $K_0$ the same as in \eqref{finite-energy}.
Therefore, we can define its Laplace-Fourier transform $w(\la,p,k)$
for any $\la>0$. 
Thanks to Theorem \ref{cor011811a}, any limit $w(\la,p,k)$
obtained this way will be constant in $k$.

In the present section we show that for any $\la>\la_0$,  see \eqref{l-0}, and $p\in\bbR$
we have  either (see \eqref{W0})
\begin{equation}
\label{diff}
\left(\la +\frac{\hat c p^2}{2}\right) w(\la,p)= \overline{ W}_0(p) ,
\end{equation}
or
\begin{equation}
\label{superdiff}
\left(\la +\hat c|p|^{3/2}\right) w(\la,p)=\overline{ W}_0(p),
\end{equation}
depending on whether  the potential is pinning or not. 
Coefficients $\hat c$ are as in Theorems \ref{diffusive-1} and
\ref{superdiffusive-1}, respectively.
Since the functions
given by either \eqref{heat1c} or \eqref{superheat1c} have the Laplace
transforms that 
satisfy \eqref{diff} and \eqref{superdiff}, respectively, we conclude
from the uniqueness of the corresponding Laplace-Fourier transform that 
 $( W_{\eps}(\cdot))$ is indeed $*-$weakly
convergent, as $\eps\to0+$, in
$L^\infty([0,T],{\cal A}')$, for any $T>0$. This would end the proof of  the assertions made
in Theorems \ref{diffusive-1} and \ref{superdiffusive-1}. The only two facts, which still require proofs are  identities (\ref{diff}) and (\ref{superdiff}).


\subsection{Derivation of (\ref{diff}) and (\ref{superdiff})}


\label{sec11.1}
Recall the definition
$$
 w_\eps^{(\iota)}(\la,p):= \int_{\bbT}  \bar w_\eps(\la,p,k) \frak
e_{\iota}(k) dk ,\qquad\iota\in\{-,+\}
$$
Let $B_M:=[p:|p|<M]$. 
In light of Theorem \ref{cor011811a} it
suffices only to show that any $*-$weak limit of
$
 w_\eps^{(+)}(\la,p)$ in $L^\infty(I\times \bar B_M)$
 satisfies  either \eqref{diff}, or \eqref{superdiff}
depending on whether  the potential is pinning or not. 

To abbreviate the notation we omit the arguments of the
functions appearing in the
ensuing calculations. From the first equation of the system \eqref{exp-wigner-eqt-1k} we get
\begin{equation}
\label{020911a}
 D^{(\eps)}\bar w_{\eps} 
=\eps^{\delta}\widehat W_{\eps}^{(0)} +\frac 32\ga\sum_{\iota\in\{-,+\}}\frak
e_{\iota} w_\eps^{(-\iota)}+q_\eps,
\end{equation}
where 
\begin{equation}
\label{030911}
 D^{(\eps)}:= \eps^{\delta} \la + 2\ga
R_\eps + i\eps\delta_\eps\om,\\
\end{equation}
and $q_\eps:=\sum_{i=1}^4 q_\eps^{(i)}$, with
\begin{eqnarray}
&&q_\eps^{(1)}
:=-
\frac{\ga (\pi\eps p)^2}{2}\left[ \frak
  f_+ \int_{\bbT} \bar w_{\eps} (4\frak f_+
 +\frak e_{-} ) dk'
 +3\frak
  f_{-} \int_{\bbT} \bar w_{\eps} \frak e_{+} dk'\right]\nonumber\\
&&
q_\eps^{(2)}:=-\ga{\cal L}\bar u_{\eps,+},\quad
q_\eps^{(3)}
:=
-i\eps\ga R'p\bar u_{\eps,-},\\
&&
q_\eps^{(4)}
:=
\frac{\ga (\pi\eps p)^2}{2} (\delta^2{\cal L}) \bar u_{\eps,+}+\eps^{3}\bar r^{(1)}_\eps.\nonumber
\end{eqnarray}
In addition, thanks to \eqref{LF1} and \eqref{C-I}, the remainder
$\bar r_\eps^{(1)}$ 
satisfies: for any compact set $I\subset (\la_0,+\infty)$ and $M>0$ 
\begin{equation}
 \label{f}
 \limsup_{\eps\to0+}\sup_{\la\in I,|p|\le M}\| r^{(1)}_\eps(\la,p)\|_{L^1(\bbT)}<+\infty.
 \end{equation}
Computing $\bar w_{\eps}$  from \eqref{020911a} and then multiplying
scalarly both sides of the resulting equation by $\ga\frak e_\iota$,
$\iota\in\{-,+\}$ we get the following system 
\begin{eqnarray*}
&&
\ga w_{\eps}^{(\iota)}\int_{\bbT}\left(1-\frac{3\ga \frak e_-\frak
    e_+}{2D^{(\eps)}}\right) dk -\frac{3\ga^2}{2}
 w_{\eps}^{(-\iota)}\int_{\bbT}\frac{ \frak e_{\iota}^2}{D^{(\eps)}} dk\\
&&
=\ga\eps^{\delta}\int_{\bbT}\frac{ \frak e_{\iota} 
\widehat W_{\eps}^{(0)} }{D^{(\eps)}} dk
+\ga\int_{\bbT} 
\frac{ \frak e_{\iota} q_\eps }{D^{(\eps)}} dk,\quad\iota\in\{-,+\}.
\end{eqnarray*}
Adding sideways the above equations corresponding to both values of
$\iota$ and then dividing both sides of the resulting equation by $\eps^{\delta}$ we obtain
\begin{equation}
\label{032301}
  a_w^{(\eps)} w_{\eps}^{(+)} -a_+^{(\eps)}\left(
    w_{\eps}^{(+)} - w_{\eps}^{(-)}\right)
  =\frac{4\ga}{3}\int_{\bbT} \frac{R\widehat
    W_{\eps}^{(0)}}{D^{(\eps)}} dk
  +\frac{4\ga}{3\eps^{\delta}}\int_{\bbT}
  \frac{R q_\eps}{D^{(\eps)}} dk,
\end{equation}
where
\begin{eqnarray}
\label{a-w}
&&
a_w^{(\eps)}(\lambda, p) :=
\frac{4\ga}{3\eps^{\delta}} \int_{\bbT}
\left(1-\frac{2\ga R}{D^{(\eps)}}\right) R \;dk\\
&&
a_+^{(\eps)}(\lambda,p):= 
\frac{\ga}{\eps^{\delta}}
\int_{\bbT}\left(1-\frac{2\ga R}{D^{(\eps)}}\right)
\frak e_+ \; dk. \nonumber
\end{eqnarray}
Let $\theta(\delta)=0$, when $\delta<2$ and $\theta(2)=1$. The following result, obviously implies
either \eqref{diff} or \eqref{superdiff}, under an  appropriate
hypothesis on the respective dispersion relation. 
\begin{prop}
\label{cor012301}
{For any $J\in {\cal S}$ such that $J(y,k)\equiv J(y)$
  and $\la>\la_0$ we have
\begin{equation}
\label{011903}
\lim_{\eps\to0+}
\left(\frac{4\ga}{3} \int_{\bbR\times \bbT} 
\frac{R\widehat W_{\eps}^{(0)}}{D^{(\eps)}}
 \hat Jdpdk- \frac{2}{3}\int_{\bbR\times \bbT} \widehat W_0 \hat J dpdk \right)=0,
\end{equation}}
In addition, for any $M>0$ and a compact interval $I\subset(\la_0,+\infty)$
\begin{equation}
\label{052301}
\lim_{\eps\to0+}\sup_{\la\in I,|p|\le M} \left|a_+^{(\eps)}(\la,p)
  \left( w_{\eps}^{(+)}(\la,p) - w_{\eps}^{(-)}(\la,p)\right)\right|=0. 
\end{equation}
Moreover, under the assumptions of Theorem \ref{diffusive-1}, we have
\begin{equation}
\label{010301}
\lim_{\eps\to0+}\sup_{\la\in I,|p|\le M}\left|a_w^{(\eps)}(\la,p)-\frac{2\la}{3}-\frac{\hat \si^2p^2}{3\ga_0}\right|=0,
\end{equation}
where $\hat \si$ is given by  \eqref{eq:sigma}, and
\begin{equation}
\label{021903}
\lim_{\eps\to0+}\sup_{\la\in I,|p|\le
  M}\left|\frac{4\ga}{3\eps^{\delta}}\int_{\bbT}\frac{R q_\eps
  }{D^{(\eps)}} dk-\frac{8\theta(\delta)\ga_0(\pi p)^2}{3}\bar
  w_\eps^{(+)}\right|=0.
\end{equation}
If, on the other hand, the assumptions of Theorem \ref{superdiffusive-1} hold then
\begin{equation}
\label{041903}
\lim_{\eps\to0+}\sup_{\la\in I,|p|\le M}\left|a_w^{(\eps)}(\la,p)-\frac{2\la}{3}-\frac{2\hat c|p|^{3/2}}{3}\right|=0,
\end{equation}
with $\hat c$ is given by \eqref{hatc}, and 
\begin{equation}
\label{051903}
\lim_{\eps\to0+}\sup_{\la\in I,|p|\le M}\left|\frac{4\ga}{3\eps^{\delta}}\int_{\bbT}\frac{R q_\eps }{D^{(\eps)}} dk\right|=0.
\end{equation}
\end{prop}
Equalities \eqref{diff} and \eqref{superdiff} then follow directly from
the above proposition by taking the limit, as $\eps\to0+$, in \eqref{032301}.

\subsection{Proof of Proposition \ref{cor012301}}


\subsection*{Proof of (\ref{011903})}
It suffices only to prove that, for any $J$ as in the
  statement of \eqref{011903}
\begin{equation}
\label{011903a}
\lim_{\eps\to0+}
\left|\frac{4\ga}{3} \int_{\bbR\times \bbT} 
\frac{R\widehat W_{\eps}^{(0)}}{D^{(\eps)}}
\hat Jdpdk- \frac{2}{3}\int_{\bbR\times \bbT} \widehat W_\eps^{(0)} \hat Jdpdk \right|=0.
\end{equation}
Then equality \eqref{011903} is a consequence of  \eqref{star-conv}. 
Note that
$2\ga R/D^{(\eps)}$ is bounded and  convergent to $1$, as
$\eps\to0+$. Using Cauchy-Schwartz inequality we can estimate the expression under the limit in  \eqref{011903a}
by
$$
\frac{2}{3}\left\{ \int_{\bbR\times \bbT} \left|\frac{2\ga R}{D^{(\eps)}}-1\right|^2 |\hat J|^2dpdk\right\}^{1/2}
\left\{\int_{\bbR\times \bbT} |\widehat W_\eps^{(0)}|^2dpdk\right\}^{1/2}.
$$
The first integral tends to $0$, as $\eps\to0+$, by virtue of the
Lebesgue dominated convergence theorem, while the second one remains
bounded thanks to condition \eqref{psi}. Thus  (\ref{011903}) follows.

\subsection*{Proof of (\ref{010301}) and (\ref{041903})}
From \eqref{a-w} we get
\begin{equation}
\label{a-w1}
a_w^{(\eps)}=\frac{4\ga}{3}
\int_{\bbT}\frac{R}{|D^{(\eps)}|^2}\left\{(\eps^{\delta}\la +2\ga R_\eps)\left(\la +\frac{\ga \eps^{2-\delta}}{4}p^2R'' \right)+\eps^{2-\delta}(\delta_\eps\om)^2\right\}dk.
\end{equation}


\subsubsection*{Diffusive scaling}

\label{diff-scale}

Here  $\om(0)>0$,  $(\delta_\eps\om)^2 \preceq R_\eps$ (see \eqref{013112a}) and, as we recall $\delta=2-s$.
Thus \eqref{010301} follows.

\bigskip

\subsubsection*{Super-diffusive scaling}

In this case $\om(k)\approx|\sin(\pi k)|$ and $\delta=(3-s)/2$. From
\eqref{a-w1} we get
\begin{equation}
\label{a-w2}
\lim_{\eps\to0+}a_w^{(\eps)}(\la,p)=\frac{2\la}{3}+\lim_{\eps\to0+}
\frac{4\ga \eps^{2-\delta}}{3}
\int_{\bbT}\frac{R(\delta_\eps\om)^2}{|D^{(\eps)}|^2}dk.
\end{equation}
We shall show that
\begin{equation}
\label{2-3-law}
\lim_{\eps\to0+}
\frac{4\ga \eps^{2-\delta}}{3}
\int_{\bbT}\frac{R(\delta_\eps\om)^2}{|D^{(\eps)}|^2}dk=\frac{2\hat c|p|^{3/2}}{3},
\end{equation}
uniformly in $\la\in I$ and $|p|\le M$. Here $\hat c$ is given by \eqref{hatc}.

 Assume that $p>0$, as
the consideration in the case $p<0$ is analogous.  Divide the domain of integration in the integral appearing in
\eqref{2-3-law} into three sets $|k|\le
\eps^{\rho_1}$, $\eps^{\rho_1}\le |k|\le \eps^{\rho_2}$ and $\eps^{\rho_2}\le |k|$ with $\rho_1>\rho_2>0$ to be adjusted later on,  and
denote the expressions corresponding to the resulting integrals by
$I_1^{(\eps)}$, $I_2^{(\eps)}$ and $I_3^{(\eps)}$.

\subsubsection*{The limit of $I_1^{(\eps)}$} Suppose that 
$
\rho_1\in (\delta-1,1)$.
Since in the unpinned case $\delta\le  3/2$,  it is
possible to find such $\rho_1$. Using the fact that
$$
|D^{(\eps)}|^2\succeq \eps |\delta_\eps\om|(\ga R_\eps+\eps^{\delta}),
$$
we conclude
$$
I_1^{(\eps)}\preceq
\eps^{1-\delta}\int_0^{\eps^{\rho_1}}\frac{\ga k^2}{\ga  k^2+\eps^{\delta}}dk
\le \eps^{1+\rho_1-\delta}\to 0,\quad\mbox{
as $\eps\to0+$.}
$$

\subsubsection*{The limit of $I_2^{(\eps)}$} 
Suppose also that $\rho_2\in(0,2-\delta-s)$. Since $s\in[0,1)$ we have
$2-\delta-s>0$ (recall that $\delta=(3-s)/2$).
We have
\begin{equation}
\label{010902-15}
\rho_2<2-\delta-s=\delta-1<\rho_1.
\end{equation}
 In  the integral
appearing in  $I_2^{(\eps)}$  we change variable according to  
$$
k':=\frac{\ga_0^{1/2}k}{C_*^{1/2}\eps^{(1-s)/2}}\quad\mbox{and}\quad C_*:=\left(\frac{\hat \al''(0)}{2}\right)^{1/2}p.
$$
Define $k_\eps:=C_*^{1/2}\eps^{(1-s)/2}k\ga_0^{-1/2}$,
$$
\tilde R(k):=\frac{\ga_0}{\eps^{1-s}C_*}R\left(k_\eps \right),\quad \tilde R_\eps(k):=\frac{\ga_0}{\eps^{1-s}C_*}R_\eps\left(k_\eps\right),
$$
and
$$
 \widehat{\delta_\eps\om}(k,p):=\frac{1}{C_*}\delta_\eps\om\left(k_\eps ,p\right).
$$
Then,
\begin{equation}
\label{010601}
I_2^{(\eps)}=\frac{(\hat
  \al''(0))^{3/4}p^{3/2}}{2^{3/4}\cdot 3\ga_0^{1/2}}\int_{{\cal
    I}_\eps}\frac{\tilde R (\widehat{\delta_\eps\om})^2 }{|\hat{D}^{(\eps)}|^2}dk,
\end{equation}
where
$$
|\hat{D}^{(\eps)}|^2:=\left(2\tilde R_\eps(k)+\frac{\eps^{\delta-1}\la}{C_*}\right)^2+ (\widehat{\delta_\eps\om})^2
$$
and
$$
{\cal I}_\eps:=\left[k:\frac{\ga_0^{1/2}\eps^{\bar\rho_1}}{C_*^{1/2}}\le
  |k|\le \frac{\ga_0^{1/2}\eps^{\bar\rho_2}}{C_*^{1/2}}\right],
$$
with $\bar\rho_i:=\rho_i-(1-s)/2$, $i=1,2$.
Note that, according to \eqref{010902-15}, 
$$
\bar \rho_1>\delta-1 - \frac{1-s}{2}=0.
$$
and
$$
\bar \rho_2<2-\delta-s - \frac{1-s}{2}=0.
$$

From \eqref{R-k} (and \eqref{beta})) we conclude that
both $\tilde R(k)$ and
$\tilde R_\eps(k)$ converge uniformly to $6\pi^2 k^2$ when $k\in {\cal
  I}_\eps$, $|p|\le M$. Likewise, $\widehat{\delta_\eps\om}(k,p)$  converges
uniformly to $1$ when $k\in {\cal
  I}_\eps$ and $\la\in I$, $|p|\le M$. Since in addition
$$
\frac{\tilde R (k) (\widehat{\delta_\eps\om}(k))^2 1_{{\cal
      I}_{\eps}}(k)}{|\hat{D}^{(\eps)}(k)|^2}
\preceq \frac{k^2+1}{k^4+1},\quad k\in\bbR
$$
we obtain
\begin{equation}
\label{020601}
\lim_{\eps\to0+}I_2^{(\eps)}
=\frac{\hat \al''(0)^{3/4}p^{3/2}}{3\cdot 2^{3/4}\ga_0^{1/2}}\int_{\bbR}\frac{24\pi^2 k^2}{144\pi^4 k^4+1}dk
\end{equation}
and the convergence is uniform in $\la\in I$ and $|p|\le M$.
Using the calculus of residua
 one can show that
 \begin{equation*}
 \label{cr-e}
\int_{-\infty}^{+\infty}\frac{k^2
  }{k^4+1}dk=\frac{\pi}{ \sqrt{2}},
\end{equation*}
therefore
$$
\int_{\bbR}\frac{24\pi^2 k^2}{144\pi^4 k^4+1}dk=\frac{1}{\sqrt{6}}.
$$
Thus,
$$
\lim_{\eps\to0+}I_2^{(\eps)}=\frac{(\hat \al''(0))^{3/4}p^{3/2}}{3^{3/2}\cdot 2^{5/4}\ga_0^{1/2}}.
$$


%
%
%

\subsubsection*{The limit of $I_3^{(\eps)}$}

Then,
$$
I_3^{(\eps)}\preceq\ga \eps^{2-\delta}
\int_{|k|\ge
  \eps^{\rho_2}}\frac{R(\delta_\eps\om)^2}{|D^{(\eps)}|^2}dk\preceq 
\ga \eps^{2-\delta}
\int_{|k|\ge
  \eps^{\rho_2}}\frac{1}{(\ga k)^2}dk\preceq
\eps^{2-\delta-s-\rho_2}\to 0,
$$
as $\eps\to0+$, uniformly in $\la\in I$ and $|p|\le M$ (recall that $\rho_2\in(0,2-\delta-s)$).
It ends the proof of \eqref{2-3-law}, thus finishing the
proof of \eqref{041903}.

\subsection*{Proof of  (\ref{052301}) }

It is a simple consequence of the following.
\begin{lemma}
\label{lm012301}
Under the assumptions of Proposition \ref{cor012301}
we have
\begin{equation*}
\label{042301c}
\sup_{\la\in I,|p|\le M}|a_+^{(\eps)}(\la,p)|\preceq 1,\quad \eps\in(0,1].
\end{equation*}
\end{lemma}
\proof
Similarly as in \eqref{a-w1} we get
$a_+^{(\eps)}=a_{+,1}^{(\eps)}+a_{+,2}^{(\eps)}$, where
\begin{eqnarray*}
&&
a_{+,1}^{(\eps)}:=\ga\int_{\bbT}\frac{\frak e_+}{|D^{(\eps)}|^2}
(\eps^{\delta}\la+2\ga
  R_\eps)\left(\la+\ga\eps^{2-\delta}\frac{p^2R''}{4}\right)dk,\\
&&
a_{+,2}^{(\eps)}:=\eps^{2-\delta}\ga\int_{\bbT}\frac{(\delta_\eps\om)^2\frak e_+}{|D^{(\eps)}|^2}
dk
\end{eqnarray*}
Term $|a_{+,1}^{(\eps)}|$  is bounded, due to the fact that
$\ga\frak e_+(\eps^{\delta}\la+2\ga
  R_\eps)\preceq |D^{(\eps)}|^2$. To bound the term 
  $a_{+,2}^{(\eps)}$ in the pinned case   we use the fact that then $(\delta_\eps\om)^2 \preceq
  R_\eps$.  In the case $\om(0)=0$ we use the bound $\frak e_+\preceq 
  R$. Then, the conclusion of the lemma follows from \eqref{2-3-law}.
\qed


\subsection*{Proof of (\ref{021903}) and (\ref{051903})}
Denote
$$
Q^{(i)}_\eps(\la,p):=\frac{4\ga}{3\eps^{\delta}}\int_{\bbT}\frac{R q_\eps^{(i)} }{D^{(\eps)}} dk.
$$
The equalities in question follow easily from our next result.
\begin{lemma}
\label{lm022301}
Under the assumptions of Proposition \ref{cor012301}
we have
\begin{equation}
\label{042301}
\lim_{\eps\to0+}\sup_{\la\in I,|p|\le M}|Q^{(i)}_\eps(\la,p)|=0,\quad i=2,3,4.
\end{equation}
Equality \eqref{042301} holds also for $Q^{(1)}_\eps(\la,p)$ when
$\delta<2$. When $\om(0)>0$ and $\delta=2$ we obtain
\begin{equation}
\label{012501}
\lim_{\eps\to0+}\sup_{\la\in I,|p|\le
  M}\left|Q^{(1)}_\eps(\la,p)-\frac{8\ga_0(\pi p)^2}{3} w_\eps^{(+)}(\la,p)\right|=0
\end{equation}
\end{lemma}
\proof
Since $|\ga R|/|D^{(\eps)}|$ is bounded the conclusion of the lemma for
  $i=4$ is a simple consequence of  \eqref{031811d}. When $i=3$, both
   $k\mapsto \delta_\eps \om(k;p)$ and $k\mapsto \hat R'(k)$ 
  are odd. Since $k\mapsto \bar u_{\eps,-}(\la,p,k)$ is even, we can
  write
$$
Q^{(3)}_\eps(\la,p)= -\frac{4\eps^{2-\delta}\ga^2}{3}
\int_{\bbT}\frac{ R R' p\; \delta_\eps\om  
\bar u_{\eps,-} }{|D^{(\eps)}|^2} dk.
$$
In case $\om(0)>0$ we use 
$
|R' p\delta_\eps\om|\preceq R_\eps
$ (see \eqref{013112a} below),
therefore
$$
|Q^{(3)}_\eps(\la,p)|\preceq \eps^{2-\delta}\int_{\bbT}|\bar u_{\eps,-}| dk
$$
and the lemma follows then by virtue of \eqref{031811d}.

In the unpinned case we use the bound
\begin{equation}
\label{D-eps}
|D^{(\eps)}|\succeq \ga R_\eps+\eps|\delta_\eps \om|
\end{equation}
together with Cauchy-Schwartz inequality and \eqref{021601}. We obtain
\begin{eqnarray*}
&&
|Q^{(3)}_\eps(\la,p)|\le \frac{4\eps^{2-\delta}\ga^2|p|}{3}\left\{\int_{\bbT}\
  R| \bar u_{\eps,-}|^2dk\right\}^{1/2}\left\{\int_{\bbT}\frac{
  R (R' \delta_\eps\om)^2  }{|D^{(\eps)}|^4}
dk\right\}^{1/2}
\\
&&
\preceq \ga^2 \eps^{2-\delta} \eps^{(\delta-s)/2}\left\{\int_{\bbT}\frac{
  R (R' \delta_\eps\om)^2  }{(\ga R_\eps)^4+\eps^4|\delta_\eps \om|^4}
dk\right\}^{1/2}
\\
&&
\le
\ga^2 \eps^{2-(\delta+s)/2} \left\{\int_{\bbT}\frac{
  R (R' \delta_\eps\om)^2  }{(\ga R_\eps)^2\eps^2|\delta_\eps
  \om|^2}
dk\right\}^{1/2}= \ga\eps^{1-(\delta+s)/2} \left\{\int_{\bbT}\frac{
  R (R')^2  }{ R_\eps^2}
dk\right\}^{1/2}\\
&&
\preceq
\eps^{(2+s-\delta)/2} \left\{\int_{0}^1\frac{
  k^3  }{ k^4+\eps^4}dk
\right\}^{1/2}\preceq
\eps^{(2+s-\delta)/2}\log^{1/2}\left(\frac{1}{\eps}\right)\to0,
\end{eqnarray*}
as $\eps\to0+$, uniformly in $\la\in I$ and $|p|\le M$.

Next, 
\begin{eqnarray}
\label{072301}
&&
Q^{(2)}_\eps(\la,p)=-\frac{4\ga}{3\eps^{\delta}}\int_{\bbT}\frac{
  R {\cal L} \bar u_{\eps,+} }{D^{(\eps)}} dk=-\frac{4\ga}{3\eps^{\delta}}\int_{\bbT}\frac{
  (R-R_\eps) {\cal L} \bar u_{\eps,+} }{D^{(\eps)}} dk\nonumber\\
&&
\\
&&
+\frac{4\ga}{3\eps^{\delta}}\int_{\bbT}\frac{
  (\eps^{\delta}\la+i\eps \delta_\eps\om) {\cal L} \bar u_{\eps,+}
}{D^{(\eps)}} dk-
\frac{4\ga}{3\eps^{\delta}}\int_{\bbT} {\cal L} \bar u_{\eps,+} dk.\nonumber
\end{eqnarray}
Denote the terms appearing on the utmost right hand of \eqref{072301}
by $I_\eps$, $I\!I_\eps$ and $I\!I\!I_\eps$, respectively.
Since $\int_\bbT{\cal L}fdk=0$ for any $f\in L^1(\bbT)$ we have
$I\!I\!I_\eps=0$. In addition, (see \eqref{022312})
 \begin{eqnarray}
\label{082301}
&&
I_\eps=\frac{\ga \eps^{\delta-2} p^2}{6}\int_{\bbT}\frac{
  R'' {\cal L} \bar u_{\eps,+} }{D^{(\eps)}}dk \\
&&
=\sum_{\iota\in\{-,+\}}\frac{\ga \eps^{\delta-2} p^2}{4} u_{\eps,+}^{(\iota)} \int_{\bbT}\frac{
  R''\frak e_{-\iota}}{D^{(\eps)}} dk -\frac{\ga\eps^{\delta-2} p^2}{3}\int_{\bbT}\frac{
  R''R }{D^{(\eps)}}\bar u_{\eps,+} dk.\nonumber
\end{eqnarray}
Here
$
 u_{\eps,+}^{(\iota)}(\la,p):=\langle \bar u_{\eps,+}(\la,p),\frak
e_{\iota}\rangle_{L^2(\bbT)}$ for $\iota\in\{-,+\}.$
Note that
\begin{equation}
\label{021501-15}
 \sup_{\eps\in(0,1]}\sup_{\la\in I,|p|\le M}\left\|\frac{
 \ga  R''R }{D^{(\eps)}}\right\|_{L^\infty(\bbT)}<+\infty.
\end{equation}
If $\om(0)>0$, then, according to \eqref{021601} we have
\begin{equation}
\label{011501-15}
G_\eps:=\int_{\bbT} 
\frac{  R_\eps|\bar u_{\eps,+}|^2 }{(\eps^{\delta}+\ga
    R_\eps)^2}dk\preceq \eps^{\delta-s},\quad \la\in I,|p|\le M,\,\eps\in(0,1].
\end{equation}
Hence, (since $\ga=\eps^s\ga_0$)
$$
| u_{\eps,+}^{(\iota)}|\preceq G_\eps^{1/2}\left\{\int_{\bbT} 
 R_\eps (\eps^{\delta}+\ga
    R_\eps)^2dk\right\}^{1/2}\preceq \eps^{(\delta+s)/2}
$$
and, using the estimate $|D_\eps|\ge \eps^{\delta}+\ga
    R_\eps(k)$,
we get
$$
\int_{\bbT}\frac{
 R  |\bar u_{\eps,+}|}{|D^{(\eps)}|} dk\preceq
G_\eps^{1/2}\preceq \eps^{(\delta-s)/2}.
$$
This leads to estimate (recall that $\delta=2-s$)
$$
|I_\eps|\preceq \eps^{\delta-2}\eps^{(\delta+s)/2}=\eps^{\delta-1}.
$$
In the unpinned case, $\bar\om(k,\eps p)\approx  R_\eps(k)$, therefore from \eqref{021601} we get
\begin{equation}
\label{011501-15a}
H_\eps:=\int_{\bbT} 
\frac{  R_\eps^2 |\bar u_{\eps,+}|^2 }{(\eps^{\delta}+\ga
    R_\eps)^2}dk\preceq \eps^{\delta-s},\quad \la\in I,|p|\le M,\,\eps\in(0,1].
\end{equation}
Hence, 
\begin{equation}
\label{041501-15}
| u_{\eps,+}^{(\iota)}|\preceq H_\eps^{1/2}\left\{\int_{\bbT} 
 (\eps^{\delta}+\ga
    R_\eps)^2dk\right\}^{1/2}\preceq \eps^{(\delta+s)/2}
\end{equation}
and, using again $|D_\eps|\ge \eps^{\delta}+\ga
    R_\eps(k)$,
we obtain
$$
\int_{\bbT}\frac{
 R  |\bar u_{\eps,+}|}{|D^{(\eps)}|} dk\preceq
H_\eps^{1/2}\preceq \eps^{(\delta-s)/2}.
$$
This leads to estimate (recall that $\delta=(3-s)/2$)
$$
|I_\eps|\preceq \eps^{\delta-2}\eps^{(\delta+s)/2}=\eps^{(1-s)/4}.
$$ We have shown therefore that in both cases $\lim_{\eps\to0+}I_\eps=0$.


Concerning term $I\!I_\eps$ note that, thanks to the fact that
$k\mapsto \delta_\eps\om(k,p)$ is odd and $k\mapsto {\cal L} \bar
u_{\eps,+}(\la,p,k)$ is even we have
\begin{eqnarray}
\label{092301}
&&
I\!I_\eps=\frac{4\ga}{3}\int_{\bbT}\frac{
  {\cal L} \bar u_{\eps,+}
}{|D^{(\eps)}|^2} \left[\la (\eps^{\delta}\la+2\ga
  R_\eps)+\eps^{2-\delta}( \delta_\eps\om)^2 \right]dk\nonumber\\
&&
\\
&&
=2\ga\sum_{\iota\in\{-,+\}}u_{\eps,+}^{-\iota}\int_{\bbT}\frac{
\frak e_{\iota}  
}{|D^{(\eps)}|^2} \left[\la (\eps^{\delta}\la+2\ga
  R_\eps)+\eps^{2-\delta}( \delta_\eps\om)^2 \right]dk\nonumber\\
&&
\nonumber\\
&&
-\frac{8\ga}{3}\int_{\bbT}\frac{
 R \bar u_{\eps,+}
}{|D^{(\eps)}|^2} \left[\la (\eps^{\delta}\la+2\ga
  R_\eps)+\eps^{2-\delta}( \delta_\eps\om)^2 \right]dk.\nonumber
\end{eqnarray}
We conclude therefore that
\begin{equation}
\label{102301}
|I\!I_\eps|\preceq \ga \eps^{2-\delta}\sum_{\iota\in\{-,+\}}| u_{\eps,+}^{-\iota}|\int_{\bbT}\frac{
\frak e_{\iota} ( \delta_\eps\om)^2
}{|D^{(\eps)}|^2} dk
+\ga \eps^{2-\delta}\int_{\bbT}\frac{
 R |\bar u_{\eps,+}|
( \delta_\eps\om)^2}{|D^{(\eps)}|^2} dk.
\end{equation}
Denote the terms appearing on the right hand side by
$I\!I_\eps^{(1)}$ and $I\!I_\eps^{(2)}$, respectively.

When $\om(0)>0$ we use the fact that $|\delta_\eps \om(k,p)|\preceq
R_\eps^{1/2}(k)$, see \eqref{013112a} below. Therefore,
$
\ga ^2R ( \delta_\eps\om)^2|D^{(\eps)}|^{-2} \preceq 1
$
and  from \eqref{031811d}, we get
$$
\lim_{\eps\to0+}\sup_{\la\in I,|p|\le M}|I\!I_\eps|=0.
$$
In the unpinned case, we use \eqref{041501-15}
  together with $|D^{(\eps)}|^2\succeq \eps|\delta_\eps \om|( \ga R_\eps)$ and get
\begin{eqnarray}
\label{112301}
&&
|I\!I_\eps^{(1)}|\preceq \ga \eps^{2-\delta}\eps^{(\delta+s)/2}\int_{\bbT}\frac{
R ( \delta_\eps\om)^2
}{|D^{(\eps)}|^2} dk\\
&&
\preceq  \ga\eps^{2+(s-\delta)/2}\int_{\bbT}\frac{
R ( \delta_\eps\om)^2
}{\eps (\ga R_\eps)|\delta_\eps\om|} dk\preceq
\eps^{(2+s-\delta)/2}\to 0,\nonumber
\end{eqnarray}
uniformly in $\la\in{\cal I}$ and $|p|\le M$, as $\delta=(3-s)/2<2+s$
for $s\in[0,1)$.

From the Cauchy-Schwartz inequality
together with \eqref{041501-15} we get
$$
|I\!I_\eps^{(2)}|\preceq \ga \eps^{2-\delta}\left\{\int_{\bbT}\frac{
 ( \delta_\eps\om)^4
}{|D^{(\eps)}|^2} dk\right\}^{1/2}\left\{\int_{\bbT}\frac{
R^2 |\bar u_{\eps,+}|^2
}{|D^{(\eps)}|^2} dk\right\}^{1/2}.
$$
We use $|D^{(\eps)}|^{2}\succeq
\eps (\eps^{\delta}+\ga R) | \delta_\eps\om|$ to estimate the first
integral and $|D^{(\eps)}|^{2}\succeq (\eps^{\delta}+\ga R)^2$
together with \eqref{011501-15a} to bound the second one. Therefore 
$$ 
\sup_{\la\in I,|p|\le M}|I\!I_\eps^{(2)}|\preceq \ga\eps^{2-\delta}\eps^{(\delta-s)/2}\left\{\int_{\bbT}\frac{
 ( \delta_\eps\om)^3
}{\eps (\ga R_\eps+\eps^{\delta})} dk\right\}^{1/2}\preceq
\eps^{(3+s)/2-\delta}\to 0,
$$
as $\eps\to0+$.
This ends the proof of
\eqref{042301}  for $i=2$.

Concerning $Q^{(1)}_\eps(\la,p)$ we can write
\begin{eqnarray*}
&&
Q^{(1)}_\eps(\la,p)
:=-\frac{2\ga^2 (\pi p)^2\eps^{2-\delta}}{3} \int_{\bbT}\frac{R }{D^{(\eps)}} 
\left[ \frak
  f_+\langle w_{\eps},4\frak f_+
+\frak e_{-}\rangle_{L^2(\bbT)}\right.\\
&&
\left.
 \qquad\qquad\qquad+3\frak
  f_{-}\langle w_{\eps},\frak
  e_{+}\rangle_{L^2(\bbT)}\right]dk.
\end{eqnarray*}
Since $\ga|R|/|D^{(\eps)}|$ is bounded the conclusion of the lemma
follows easily for $\delta<2$. If  $\delta=2$, then $\ga\equiv \ga_0$. We can use the
Lebesgue dominated convergence theorem and obtain
\eqref{012501}.
\qed



\subsection{The dual dynamics}

\label{sec11.3}

The equations  \eqref{exp-wigner-eqt-1} and \eqref{anti-wigner-eqt1}
describing the dynamics of the column vector $\widehat{\mathbb W}_\ep
(t,p,k)$ given by
$$
\widehat{\mathbb W}_\ep^T (t,p,k) = [\widehat
W_{\ep,+}(t,p,k), \widehat Y_{\ep,+}(t,p,k), \widehat Y_{\ep,-}(t,p,k),
\widehat W_{\ep,-}(t,p,k)]
$$ 
can be written in  the form
\begin{equation}
\label{dyn-vect}
\frac{d}{dt}\widehat{\mathbb W}_\ep (t,p,k) =
\mathbb L_\eps \widehat{\mathbb W}_\ep (t,p,k) 
\end{equation}
where $\mathbb L_\eps$ is some  matrix operator.
We now define the dual dynamics that  runs on  test functions. Suppose that 
$$
(\widehat{\mathbb J}^{(\eps)})^T(t,p,k)
=\left[
\widehat J^{w,+}_\eps(t),
\widehat J^{y,+}_\eps(t),
\widehat J^{y,-}_\eps(t),
\widehat J^{w,-}_\eps(t)
\right],
$$
is the solution of the system dual to
\eqref{dyn-vect}, i.e. 
\begin{equation}
\label{033001}
\frac{d}{dt}\widehat{\mathbb J}^{(\eps)}(t)= 
\mathbb L_\eps^*\widehat{\mathbb J}^{(\eps)}(t). 
\end{equation}
with given initial conditions
 that are the Fourier transforms of some functions belonging to ${\cal S}$.
The adjoint matrix ${\mathbb L}_\eps^*$ is given explicitely by
\begin{equation}
\label{L-eps}
{\mathbb L}_\eps^*\widehat{\mathbb J}
:=\left[
\begin{array}{llll}
L_w^{(\eps)} &L_-^{(\eps)} &L_+^{(\eps)} &0\\
L_-^{(\eps)} &L_y^{(\eps)}&R_y^{(\eps)} &L_+^{(\eps)} \\
L_-^{(\eps)} &R_y^{(\eps)} &\bar L_y^{(\eps)}&L_+^{(\eps)} \\
0&L_-^{(\eps)} &L_+^{(\eps)} &\bar L_w^{(\eps)}
\end{array}
\right]\widehat{ \mathbb J},
\end{equation}
with
\begin{eqnarray*}
&&
L_w^{(\eps)}:=\eps^{-\delta}(i\eps \delta_\eps \om(k,p)+ \ga {\cal
  L}^*_{\eps p}),\qquad 
\bar L_w^{(\eps)}:=\eps^{-\delta}(-i\eps \delta_\eps \om(k,p) +\ga {\cal L}^*_{\eps p}),\\
&&
L_y^{(\eps)}:=\eps^{-\delta}[i \bar \om(k,\eps p)+\ga
({\cal L}^*_{\eps p} - {\cal R}^*_{\eps p})],\qquad 
\bar L_y^{(\eps)}:=\eps^{-\delta}[-i \bar \om(k,\eps p)+\ga
({\cal L}^*_{\eps p} - {\cal R}^*_{\eps p})],\\
&&
L_\pm^{(\eps)}:=-2^{-1}\ga\eps^{-\delta} \left({\cal L}^\pm_{\eps p}\right)^*,
\qquad R_y^{(\eps)} :=\eps^{-\delta}\ga {\cal R}^*_{\eps p}.
\end{eqnarray*}
The operators ${\cal L}^*_p$, $\left({\cal L}^\pm_{p}\right)^*$ and 
 ${\cal R}^*_p$ are the
 adjoints of ${\cal L}_p$, ${\cal L}^\pm_{p}$ and 
 ${\cal R}^*_p$ (see \eqref{010511} and \eqref{050411}) with respect to the Lebesgue measure on $\bbT$.
%
Given $M>0$ we introduce the norm
\begin{equation}
\label{norm-ta1}
\| J\|_{ {\cal A}_{2,M}}:=\int_{B_M}dp
  \left\{\int_{\bbT}|\hat J(p,k)|^{2}dk\right\}^{1/2}, 
\end{equation}
and  denote by ${\cal A}_{2,M}$  the completion of ${\cal S}$ under the  norm and by ${\cal A}_{2,M}'$ its dual, that is the space of all $J:\bbR\times \bbT\to\mathbb C$ equipped with the norm
\begin{equation}
\label{norm-tainf}
\| J\|_{ {\cal A}_{2,M}'}:=\sup_{|p|\le M}
  \left\{\int_{\bbT}|\hat J(p,k)|^{2}dk\right\}^{1/2}.
\end{equation}
Let ${\cal A}_{2,loc}:=\bigcap_{M>0}{\cal A}_{2,M}$. 
Given $J\in{\cal S}$, define
$$
 \underbar{J}(p):= \int_{\bbT} \hat J(p,k) dk =
 \int_{\bbR\times \bbT}e^{-2\pi i py}J(y,k)dydk
$$
and
$$
\varphi(p):=\left\{
\begin{array}{ll}
 \dfrac{\hat c p^2}{2},&\mbox{ in case $\hat \al(0)>0$,}\\
&\\
\hat c |p|^{3/2} ,&\mbox{ in case $\hat \al(0)=0$.}
\end{array}
\right.
$$
Coefficient $\hat c$ is determined either as  in Theorem
\ref{diffusive-1}, when $\hat\al(0)>0$, or 
Theorem \ref{superdiffusive-1}, when $\hat\al(0)=0$.
We can repeat the argument made so far and  conclude the following
statement concerning the convergence of the dual dynamics. 
\begin{prop}
\label{dual-thm}
Suppose that the initial data in  \eqref{033001}
belongs to ${\cal A}_{2,loc}$. Then the
following are true: 
\begin{itemize}
\item[i)]
 for any $T,M>0$ we have
$$
\sup_{\eps\in(0,1]}\sup_{t\in[0,T]}\|\mathbb J^{(\eps)}(t)\|_{{\cal A}_{2,M}'}<+\infty,
$$ 
\item[ii)] suppose that $M>0$ is such that $\widehat{ \mathbb J}^{(\eps)}(0,p,k)\equiv0$ for $|p|\ge M$ and $k\in\bbT$. Then
$\widehat{ \mathbb J}^{(\eps)}(t,p,k)\equiv0$ for $t\ge0$, $|p|\ge M$ and $k\in\bbT$,
\item[iii)]
for any $T>0$ and $g\in L^1([0,T],{\cal A}_{2,loc})$ and $\widehat{
  \mathbb J}^{(\eps)}(t,p,k)$ as in ii)  we have
\begin{eqnarray}
\label{020402}
&&\lim_{\eps\to0}\int_0^T\int_{B_M\times \bbT} 
\hat J^{w,+}_\eps(t,p,k)\hat g^*(t,p,k)dtdpdk\nonumber\\ 
&& \qquad
=\int_0^T\int_{\bbR}e^{-\varphi(p) t}\underbar J^{w,+}(p){\underbar g}^*(t,p)dtdp.
\end{eqnarray}
\end{itemize}
\end{prop}

\section{Proofs of Theorems \ref{diffusive-1a} and \ref{superdiffusive-1a}}

\label{diff-scale-1}

\subsection{Evolution of the random Wigner transform}

%

To describe the evolution of the fluctuating Wigner transform
$\widetilde{\cal W}_\eps(t;J)$, see \eqref{wigner-def1}, we shall
also need the following quantities 
 \begin{eqnarray*}
&&
\widetilde{\cal Y}_{\eps,+}(t;J):=\sqrt\eps \sum_{x,x'\in\bbZ}  \psi^{(\eps)}_{x'}(t)
  \psi_x^{(\eps)}(t) 
  \tilde J^*\left(\frac{\eps}{2}(x+x'), x'-x\right),\nonumber\\ 
&&
\\
&&
\widetilde{\cal Y}_{\eps,-}(t;J):=\sqrt\eps\sum_{x,x'\in\bbZ}
\left(\psi^{(\eps)}_{x'}(t) 
  \psi_x^{(\eps)}(t)\right)^* 
\tilde J^*\left(-\frac{\eps}{2}(x+x'), x'-x\right),\nonumber\\ 
  &&
  \nonumber\\
&&
\widetilde{\cal   W}_{\eps,-}(t;J):= \sqrt\eps \sum_{x,x'\in\bbZ}
\left(\psi^{(\eps)}_{x'}(t) 
\left(  \psi_x^{(\eps)}(t)\right)^*- \delta_{x,x'} \mathcal E_0\right)
 \tilde J^*\left(-\frac{\eps}{2}(x+x'),x'-x\right).\nonumber 
\end{eqnarray*}
We  identify   $\widetilde{\cal W}_{\eps,+}(t)=\widetilde {\cal
  W}_{\eps}(t)$.
Given the column vector $\mathbb J^T= [J^{w,+},J^{y,+},J^{y,-}, J^{w,-}]$ with the
components that are the Fourier transforms of functions from $\cal S$, denote 
$$
\widetilde{\mathbb W}_\eps(t;\mathbb J)
:=
\sum_{\iota\in\{-,+\}}
\left(\widetilde{\cal W}_{\eps,\iota}(t;J^{w,\iota})+
\widetilde{\cal Y}_{\eps,\iota}(t;J^{y,\iota})\right).
$$
For  $J_1,J_2\in{\cal S}$ let
  \begin{eqnarray}
\label{wigner-cov1}
&&
C_\eps^{w,\pm}(t;J_1,J_2):= \bbE\left[\widetilde{\cal W}_{\eps,\pm}(t;J_1)
\widetilde {\cal W}_\eps(0;J_2)\right],\nonumber\\
&&
\\
&&
C_\eps^{y,\pm}(t;J_1,J_2):= \bbE\left[\widetilde{\cal Y}_{\eps,\pm}(t;J_1)
 \widetilde{\cal W}_\eps(0;J_2)\right].\nonumber
\end{eqnarray}

Computing the time differential as in Section \ref{sec5}, we obtain
 \begin{eqnarray}
\label{exp-wigner-eqt-xx}
&&
d  \widetilde{\cal W}_\eps(t;J) = \eps^{-\delta} 
\left\{\widetilde{\cal W}_\eps 
\left(t;\left(i\eps A+ \ga
{\cal L}^*_{\eps}\right)J\right) -\frac{\ga}{2}
\sum_{\si\in\{-,+\}}\widetilde{\cal Y}_{\eps,\si}(t;{\cal L}^*_{\eps,-\si} J)
\right.\nonumber\\ 
&&
\\
&&
 \left. + \eps^{-1/2} {\cal E}_0\sum_{n\in\bbZ}\int_{\bbT}  \left(
i\eps A \hat J+ \ga
{\cal L}^*_{\eps}\hat J\right) \left(\frac{n}{\eps},k\right)dk \right\}dt+d{\cal M}^{(\eps)}_t(J),\nonumber
\end{eqnarray}
\begin{equation}
\label{exp-wigner-eqt-xy}
\begin{split}
 & d\widetilde{\cal Y}_\eps(t;J)= \eps^{-\delta}\Big\{\widetilde{\cal
      Y}_\eps \left(t;\left(i B + \ga{\cal L}^*_{\eps}\right)J\right)
  + \ga\widetilde{\cal Y}_{\eps,-}\left(t;{\cal R}^*_{\eps}J\right) -\ga
    \widetilde{\cal Y}_{\eps}\left(t;{\cal R}^*_{\eps}J\right) 
  \\
&  -\frac{\ga}{2} \sum_{\si\in\{-,+\}}
    \widetilde{\cal W}_{\eps,\si}(t;{\cal L}^*_{\eps,-\si} J)
   -\frac{\ga {\cal E}_0}{2 \eps^{1/2}} \sum_{n\in\bbZ}\int_{\bbT}  
\left({\cal L}^*_{\eps,-\si} \hat J\right) \left(\frac{n}{\eps},k\right)dk
  \Big\}dt +d{\cal N}^{(\eps)}_t(J) ,
\end{split}
\end{equation}
where  ${\cal M}^{(\eps)}_t(J),$  ${\cal N}^{(\eps)}_t(J),$  are some square integrable, continuous trajectory martingales.
Summarizing, if  the test functions $J^{w,\pm}$ and $J^{y,\pm}$ are such that
their respective Fourier transforms in the $x$ variable  $\hat J^{w,\pm}$
and $\hat J^{y,\pm}$ belong to $C_c^\infty(\bbR\times\bbT)$, then,
using \eqref{exp-wigner-eqt-xx} and \eqref{exp-wigner-eqt-xy}, we obtain
\begin{equation}
\label{023001}
 \frac{d}{dt}\bbE\left[\widetilde{\mathbb W_\eps}(t;\mathbb J)
\widetilde{\cal W}_\eps(0;J)\right]
=\bbE\left[\widetilde{\mathbb W_\eps}(t;\mathbb L_\eps^*\mathbb J)
\widetilde{\cal W}_\eps(0;J)\right] 
\end{equation}
where $\mathbb L_\eps^*$ is given by \eqref{L-eps}.

Suppose that 
$
\mathbb J^{(\eps)}(t)$
is the solution of the equation \eqref{033001}.
From part ii) of Proposition \ref{dual-thm} we conclude that
\begin{equation}
\label{support}
\,\widehat{\mathbb J}^{(\eps)}(t,p,k)\equiv0,\quad \forall \,t\ge0,\,|p|\ge M,\,k\in\bbT,
\end{equation}
provided that $M>0$ is such that  $\widehat{\mathbb
  J}^{(\eps)}(0,p,k)\equiv0$ for all $|p|\ge M$,  $k\in\bbT$.
Combining \eqref{023001} with \eqref{033001} we
obtain
\begin{eqnarray*}
&& \frac{d}{ds}\bbE\left[\widetilde{\mathbb W}_\eps(s;\mathbb J^{(\eps)}(t-s))
\widetilde{\cal  W}_\eps(0;J)\right]=
\bbE\left[\widetilde{\mathbb W}_\eps(s;\mathbb L_\eps^*\mathbb J^{(\eps)}(t-s)) 
\widetilde{\cal W}_\eps(0;J)\right]\nonumber\\
&&
\\
&&
+\bbE\left[\widetilde{\mathbb W}_\eps\left(s;\frac{d}{ds}\mathbb
    J^{(\eps)}(t-s)\right) \widetilde{\cal W}_\eps(0;J)\right]
\equiv 0\nonumber
\end{eqnarray*}
for all $s\ge0$. Comparing the values of 
$\bbE\left[\widetilde{\mathbb W}_\eps(s;\mathbb J^{(\eps)}(t-s))\widetilde{\cal  W}_\eps(0;J)\right]$ for
  $s=t$ and $s=0$ we get
\begin{equation}
\label{043001}
\bbE\left[\widetilde{\mathbb W_\eps}(t;\mathbb J)
\widetilde {\cal  W}_\eps(0;J)\right] =
\bbE\left[\widetilde{\mathbb W}_\eps(0;\mathbb J^{(\eps)}(t)) 
 \widetilde{\cal W} _\eps (0;J)\right],
\end{equation}
where
$$
\mathbb J:=\mathbb J^{(\eps)}(0)=[J^{w,+},J^{y,+},J^{y,-},J^{w,-}]^T.
$$
Suppose that the initial data satisfies the hypothesis of part ii) of
Proposition \ref{dual-thm} and that $\hat J$ is compactly supported. According to
\eqref{covariance} the right hand side of \eqref{043001} equals
 \begin{eqnarray*}
 &&
 \bbE\left[\widetilde{\cal W}_\eps(0;J^{w,+}_\eps(t))
  \widetilde{\cal W} _\eps (0;J)\right]
= {\cal E}_0^2\eps\sum_x \int_{\bbT} J^{w,+}_\eps\left(t,\eps
    x,k\right)   J^*\left(\eps x,k\right) dk\\
&&
= {\cal E}_0^2\sum_n \int_{\bbR\times\bbT} \hat
J^{w,+}_\eps\left(t,\frac{n}{\eps}-p,k\right) \hat  J^*\left(-p,k\right)
dpdk
\end{eqnarray*}
The last equality holds, thanks to the Poisson summation formula, see
\cite{lax}, formula (50)  on p. 566.
Since the supports of $\hat
J^{w,+}_\eps\left(t\right)$ and $ \hat J$ are both compact in $p$ for a
sufficiently small $\eps$ we can write that the right hand side equals
$$
 {\cal E}_0^2 \int_{\bbR\times\bbT} 
\hat J^{w,+}_\eps\left(t,p,k\right) \hat J^*\left(p,k\right) dpdk. 
$$
Using \eqref{020402} we conclude that for any compactly supported
$\phi \in L^1[0,+\infty)$ 
we have
\begin{eqnarray*}
&&
\lim_{\eps\to0+}\int_0^{+\infty}\phi(t)\bbE\left[\widetilde{\cal W}_\eps(t;J^{w,+})
 \widetilde{\cal  W}_\eps(0;J)\right]dt\\
&&
=\lim_{\eps\to0+}\int_0^{+\infty}\phi(t)\bbE\left[\widetilde{\cal
    W}_\eps(0; J^{w,+}_\eps(t)) 
  \widetilde{\cal W} _\eps (0;J)\right]dt\\
&&
={\cal E}_0^2 \int_0^{+\infty}dt\int_{\bbR}\phi(t)\exp\left\{-\frac{\hat
    c p^2 t}{2}\right\}\underbar J^{w,+}(p)\underbar J(p)dp,
\end{eqnarray*}
with $\hat c$ given by  \eqref{hatc} when $\delta<2$, or \eqref{hatc1}
when $\delta=2$ in the case of a pinning potential, or 
\begin{eqnarray*}
&&
\lim_{\eps\to0+}\int_0^{+\infty}\phi(t)\bbE\left[\widetilde{\cal W}_\eps(t; J^{w,+})
 \widetilde{\cal  W}_\eps(0;J)\right]dt\\
&&
={\cal E}_0^2 \int_0^{+\infty}dt\int_{\bbR}\phi(t)\exp\left\{-\hat
    c |p|^{3/2} t\right\}\underbar  J^{w,+}(p)\underbar J(p)dp,
\end{eqnarray*}
with $\hat c$ given by \eqref{hatc-32} in the unpinned
case. Generalization to arbitrary $ J^{w,+},J\in {\cal S}$ and $\phi\in
L^1[0,+\infty)$ is standard and can be done via  an approximation, due
to the fact that process
$  \left(\widetilde{\cal W} _\eps (t;J)\right)_{t\ge0}$ is stationary.

\textbf{Remark:} Observe that the proof does not really use time
stationarity of the initial distribution, in fact it follows that for
any initial homogeneous distribution with energy density given by some
$\cal E(k)$ such that $\int_{\bbT} \cal E(k) dk = 2{\cal E}_0$, we have the same
result. On the other hand, we do use the stationarity in order to prove the
equivalence of the energy distribution \eqref{011704}, see Section \ref{sec:equiv-betw-en}.

\section{Equivalence of energy functionals}

\label{sec:equiv-betw-en}

\subsection{Proof of Proposition \ref{prop011404}}

\label{sec:equiv-betw-wign}

\subsubsection{The case of a pinned potential}

The left hand side of \eqref{011803} equals $\lim_{\eps\to0+}(I_\eps+I\! I_\eps)$, where
\begin{eqnarray*}
&&
I_\eps:=\frac{\eps}{4}\sum_{x}J(\eps x)  \bbE_{\eps}\left(\hat\al(0)[{\frak q}_x^{(\eps)}(t)]^2-\sum_y\al_{x-y}[{\frak q}_y^{(\eps)}(t)]^2\right)\\
&&
=
\frac{\eps}{4}\sum_{x}\bbE_{\eps}[{\frak q}_x^{(\eps)}(t)]^2\int_{\bbR}e^{2\pi i\eps xp}\hat J(p)  [\hat\al(0)-\hat\al(-\eps p)]dp
\end{eqnarray*}
and
\begin{eqnarray*}
&&
I\!I_\eps:=\frac{\eps}{2}\sum_{x}J(\eps x)  \bbE_{\eps}\left[{\frak q}_x^{(\eps)}(t)(\al*{\frak q}^{(\eps)}(t))_x-(\tilde\om*{\frak q}^{(\eps)}(t))_x^2\right]\\
&&
=
\frac{\eps}{2}\sum_{y,y'}\bbE_{\eps}({\frak q}_y^{(\eps)}(t){\frak q}^{(\eps)}_{y'}(t))\sum_x\int_{\bbR}e^{2\pi i\eps xp}\hat J(p)  [\tilde \om_{x-y}(0)-\tilde \om_{x-y}(-\eps p)]\tilde \om_{x-y'}dp.
\end{eqnarray*}
Here we have adopted the notation
$
\tilde \om_{x}(p):=e^{-2\pi i  xp}\tilde \om_x. 
$
Since in the case of a pinned chain  we have
$$
\sup_{\eps\in(0,1]}\eps\bbE_{\eps}[{\frak q}_x^{(\eps)}(t)]^2<+\infty
$$
it is clear  that
for any $J\in C_0^\infty(\bbR)$ we have
\begin{equation}
\label{021803}
\lim_{\eps\to0+} I_\eps=0\qquad \mbox{ and} \qquad \lim_{\eps\to0+} I\!I_\eps=0.
\end{equation}


\subsubsection{The case of an unpinned potential}

Then $\hat\al(0)=0$. Let $(\mu_\eps)_{\eps\in(0,1]}$, be the family of  probability distributions on $\ell^2$ such that
condition \eqref{psi} holds.
We claim that it suffices to prove  \eqref{011803}  only for $t=0$.
Indeed, our argument, presented below, shows that the
   equivalence of energy density functionals 
  is a consequence of  the aforementioned bound \eqref{psi}.
By virtue of 
the estimate \eqref{021501} for $p=0$ this bound persists in time, so our proof
shows that in fact \eqref{011803} holds for  any  subsequent  time
$t\ge0$. 
For the purpose of this proof we let 
\begin{equation}
\label{011405}
{\frak q}_x:=\int_{\bbT}(e^{2\pi i kx}-1)\hat {\frak q}(k)dk,
\end{equation}
where
$$
\hat {\frak q}(k):=\frac{\hat\psi(k)+\hat\psi^*(-k)}{2\om(k)}.
$$
 As a consequence of \eqref{finite-energy} and \eqref{psi} we obtain
 \begin{equation}
  \label{eq:finiteenergy}
  \limsup_{\eps\to0+} \eps\left\langle \sum_x(\tilde\om*{\frak q})_x^2 \right\rangle_{\mu_\eps} < +\infty
\end{equation}
and
\begin{equation}\label{eq:l2energy}
 \limsup_{\eps\to0+} \int_\bbT \left[\eps\left<|\om(k)\hat {\frak q}(k)|^2
        \right>_{\mu_\eps}\right]^2dk\;  <+\infty.
\end{equation}

 Define 
\begin{equation}
\label{012304}
\delta\hat\al(k,k'):=\hat \al(k+k')-\hat \al(k)-\hat \al(k').
\end{equation}
\begin{lm}
\label{lm012204}
We have
\begin{equation}
\label{022204}
\left|\frac{\delta\hat \al(k,k')}{\om(k)\om(k')}\right|\preceq 1,\quad k,k'\in \bbT.
\end{equation}
\end{lm}
\proof
Observe that 
$$
\hat \al(k)=-2\sum_x\al_x{\frak s}^2(xk).
$$
Therefore \eqref{022204} is a consequence of the following elementary
inequality
$$
|{\frak s}^2(\al+\bt) -{\frak s}^2(\al)-{\frak s}^2(\bt)|\preceq
|{\frak s}(\al) {\frak s}(\bt)|,\quad \al,\beta\in\bbR
$$
and the assumption a1) made on the decay of $(\al_x)$. 
\qed

 Let
\begin{equation}
  \label{eq:phi}
  \phi_x := |\psi_x|^2-2{\frak e}_x=\frac 12\sum_y \alpha_{x-y} ({\frak q}_x -{\frak q}_y)^2 +
  (\tilde\omega*{\frak q})_x^2 
\end{equation}
Obviously Proposition \ref{prop011404}  is a consequence of the following.
\begin{lemma}\label{lem-simp}
  For any $J\in C_0^\infty(\R)$
  \begin{equation}
    \label{eq:equine}
    \lim_{\eps\to 0} \eps\sum_x J(\eps x) \left< \phi_x
    \right>_{\mu_\eps}\ =\ 0.
  \end{equation}
\end{lemma}
\proof
 Using  \eqref{011405} we can write $\phi_x$ in Fourier transform coordinates as
  \begin{equation}
    \label{eq:ftphi}
    \phi_x = \iint_{\bbT^2}  \hat F(k,k')
    \omega(k)\hat{\frak q}(k) \omega(k')\hat{\frak  q}(k') e^{i2\pi (k+k')x}dk\; dk'\;
  \end{equation}
where
\begin{equation}
  \label{012304b}
  \hat F(k,k'):=\frac{\hat \al(k+k')-\hat \al(k)-\hat
  \al(k')}{2\omega(k)\omega(k')} + 1.
\end{equation}
Note that $F(k,-k)=0$. Moreover, according to Lemma \ref{lm012204} it is bounded.
Observe that under the condition \eqref{eq:finiteenergy}, function $\om(k)\hat {\frak q}(k)$ is square integrable on $\bbT$ (although $\hat {\frak q}(k)$  need not be so).
Furthermore,
$$
    \eps \sum_x  J(\eps x) \left<\phi_x \right>_{\mu_\eps} =\  \int_{\R}
    \hat J(p) Z_\eps(p) dp,
 $$
     where
   $$
   Z_\eps(p) :=  \eps\int_{\bbT} \hat F(k,-k-\eps p) \left<\omega(k)\hat{\frak q}(k)
      \omega(-k-\eps p) \hat{\frak  q}(-k-\eps p) \right>_{\mu_\eps}
    \;dp\; dk .
   $$
By the Schwarz inequality and  symmetryin $k$ and $k+\eps p$, we obtain
\begin{equation*}
  \begin{split}
    |Z_\eps(p)| \le \int_\bbT dk\; |\hat F(k,-k-\eps p)|
    \eps\left<\omega(k)^2|\hat {\frak q}(k)|^2 \right>_{\mu_\eps}\\
    \le \left(\int_\bbT dk\; |\hat F(k,-k-\eps p)|^2\right)^{1/2} 
    \left(\int_\bbT dk\; \left[\eps\left<\omega(k)^2|\hat{\frak  q}(k)|^2
        \right>_{\mu_\eps}\right]^2 \right)^{1/2}\\
    \le C \left(\int_\bbT dk\; |\hat F(k,-k-\eps p)|^2\right)^{1/2} 
    \mathop{\longrightarrow}_{\eps\to 0+} 0.
  \end{split}
\end{equation*}
Since $Z_\eps(p)$ is bounded, the result follows upon an application of the Lebesgue dominated convergence theorem.
 \qed


\subsection{Proof of Proposition \ref{prop012204}}

\label{sec13.3}

 Obviously, stationarity implies that the limit in \eqref{011704} does not depend on
  $t$ therefore it suffices to prove  \eqref{011704} for $t=0$. For
  that purpose it is
  enough to 
  show that
\begin{equation}
\label{021704}
\lim_{\eps\to0+}\eps \bbE\left[\sum_x\phi_xJ(\eps x)\right]^2=0,
\end{equation}
where $\phi_x=\sum_{i=1}^3 \phi_x^{(i)} $ and 
\begin{eqnarray*}
&&
\phi_x^{(1)} := \frac 12 \sum_{x'} \alpha_{x-x'} (\frak q_x
 -\frak q_{x'})^2,\\
&&    \phi_x^{(2)} := (\tilde\omega*\frak q)_x^2,\qquad
\phi_x^{(3)} :=-\hat\al(0)\frak q_x^2.
\end{eqnarray*}


\subsubsection{The case of an unpinned chain}
We assume that $\hat\al(0)=0$, therefore $\phi_x^{(3)} =0$.  
Then, the field $({\frak q}_x)$ is  Gaussian given by
\begin{equation}
\label{021904}
\frak q_x  =\sqrt{{\cal
  E}_0}\,\int_{\bbT} \frac{e^{2\pi i kx}-1}{\om(k)}\hat w(dk),
\end{equation}
where $\hat w(dk)$ is a complex even, Gaussian white noise in $L^2(\bbT)$, i.e. 
$$
\bbE[\hat w(dk) \hat w^*(dk')]=\delta(k-k')dkdk',\quad \hat w^*(dk) =\hat w(-dk).
$$
As a result
\begin{equation}
\label{011804}
\bbE(\frak q_x  \frak q_{x'})={\cal
  E}_0\int_{\bbT}[e^{2\pi i kx}-1][e^{2\pi i kx'}-1]^*\al^{-1}(k)dk
\end{equation}
Note that $\bbE\phi_x= 0$ for all $x\in\bbZ$. Indeed, we have
\begin{eqnarray*}
&&
\bbE\phi_x^{(1)} = \frac 12 \sum_{x'} \alpha_{x-x'}\bbE (\frak q_x
 -\frak q_{x'})^2
 = \frac 12 \sum_{x'} \alpha_{x'}\bbE \frak q_{x'}^2
\\
&&
=\frac{{\cal
  E}_0}{2}\sum_{x'}\al_{x'}\int_{\bbT} [2-(e^{2\pi i kx'}+e^{-2\pi i kx'}
)]\al^{-1}(k)dk
=-{\cal
  E}_0\int_{\bbT} \frac{ \al(k) }{\al(k)}dk=-{\cal
  E}_0.
\end{eqnarray*}
In addition, thanks to \eqref{011804},
\begin{eqnarray*}
&&
\bbE\phi_x^{(2)} = \sum_{x',x''} \tilde \om_{x-x'}\tilde\om_{x-x''}\bbE(\frak q_{x'}
 \frak q_{x''})
\\
&&
={\cal
  E}_0\int_{\bbT}\left(\sum_{x',x''} \tilde
\om_{x-x'}\tilde\om_{x-x''}e^{2\pi i k(x'-x'')} \right)\al^{-1}(k)dk={\cal
  E}_0\int_{\bbT}\om^2(k)\al^{-1}(k)dk={\cal
  E}_0.
\end{eqnarray*}
In our next step we calculate
\begin{equation}
\label{011904}
r_x:=\bbE(\phi_x \phi_0)= \sum_{i,i'=1}^2 r_x^{(i,i')},
\end{equation}
where
$
r_x^{(i,i')}:=\bbE(\tilde\phi_x^{(i)}\tilde \phi_0^{(i')})
$
and
$
\tilde \phi_x^{(1)}:=\phi_x^{(1)}+{\cal
  E}_0$, and $ \phi_x^{(2)}:=\phi_x^{(2)}-{\cal
  E}_0.
$
We have
\begin{eqnarray*}
 r_x^{(1,1)}=&&\frac12\sum_{x',x''} \alpha_{x-x'}\alpha_{x''}\left\{\bbE[ (\frak q_x
 -\frak q_{x'}) 
 \frak q_{x''} ]\right\}^2\\
&&=\frac{{\cal E}_0^2}{2}\int_{\bbT^2}[\delta\hat \al(k,k')]^2 
\frac{ e^{2\pi i (k+k')x}}{\hat\al(k)\hat \al(k')}dkdk',
\end{eqnarray*}
\begin{eqnarray*}
 r_{x}^{(2,1)}= r_x^{(1,2)}= &&\sum_{x'} \alpha_{x-x'}\left\{\bbE[ (\frak q_x
 -\frak q_{x'}) (\tilde\om*{\frak q})_0]\right\}^2 \\
&&={\cal E}_0^2\int_{\bbT^2}\delta\hat \al(k,k')
\frac{e^{2\pi i(k+k')x}}{\om(k)\om(k')}dkdk',
\end{eqnarray*}
$$
r_x^{(2,2)}= 2\left\{\bbE[ 
 (\tilde\om*{\frak q})_0 (\tilde\om*{\frak q})_x]\right\}^2
=2{\cal E}_0^2\delta_{x,0}.
$$
Here $\delta\hat\al(k,k')$ is given by \eqref{012304}.
Therefore,
\begin{eqnarray}
\label{012204}
&&
\eps \bbE\left[\sum_x\phi_xJ(\eps x)\right]^2 =\eps \sum_{x,x'}J(\eps x) J(\eps x')r_{x-x'}\\
&&
=\frac{\eps{\cal E}_0^2}{2}\sum_{x,x'}\int_{\bbR^2\times\bbT^2}e^{2\pi i(k+k')(x-x')}e^{2\eps\pi i(xp+x'p')}\hat J(p)\hat J(p')F(k,k')dpdp'dkdk',\nonumber
\end{eqnarray}
with
$$
F(k,k'):=\frac{[\delta\hat \al(k,k')+2\om(k)\om(k')]^2}{\hat\al(k)\hat
    \al(k')}.
$$
Observe that
$F(-k,k)=0$.
Summing first over $x$ and then over $x'$ we obtain that the utmost
right hand side of \eqref{012204} equals
\begin{equation}
\label{012204a}
\sum_{n\in\bbZ}\int_{\bbR}\hat J(p)\hat
J\left(\frac{n}{\eps}-p\right)\left(\int_{\bbT} F(-k-\eps p,k)dk\right)dp.
\end{equation}
Therefore \eqref{021704} (thus also the conclusion of the proposition) is a
consequence of the Lebesgue dominared convergence theorem and
Lemma \ref{lm012204}.

\bigskip

\subsubsection{The pinned case}
Then,
\begin{equation}
\label{062204}
\frak q_x  =\sqrt{{\cal
  E}_0}\,\int_{\bbT}\frac{e^{2\pi i kx} }{\om(k)}\hat w(dk).
\end{equation}
We have
$\bbE\phi_x=0$ with
\begin{eqnarray*}
&&
 \bbE\phi_x^{(1)}={\cal
  E}_0\int_{\bbT} \frac{\hat \al(0)-\hat\al(k)}{\hat\al(k)}dk,\\
 &&
\bbE\phi_x^{(2)}={\cal
  E}_0,\qquad
\bbE\phi_x^{(3)}=-{\cal
  E}_0\int_{\bbT} \frac{\hat \al(0)}{\hat\al(k)}dk.
\end{eqnarray*}
We let
\begin{equation}
\label{011904a}
r_x:=\bbE(\phi_x \phi_0)= \sum_{i,i'=1}^3 r_x^{(i,i')},
\end{equation}
where
$
r_x^{(i,i')}:=\bbE(\tilde\phi_x^{(i)}\tilde \phi_0^{(i')})
$
and
$
\tilde \phi_x^{(i)}:=\phi_x^{(i)}-\bbE\phi_x^{(i)}$, $i=1,2,3.$
We have
\begin{eqnarray*}
&& r_x^{(1,1)}=\frac12\sum_{x',x''} \alpha_{x-x'}\alpha_{x''}\left\{\bbE[ (\frak q_x
 -\frak q_{x'}) (\frak q_{x''}-\frak q_0)]\right\}^2\\
&&
=\frac{{\cal E}_0^2}{2}\int_{\bbT^2}[\delta\hat \al(k,k')]^2\frac{e^{2\pi i(k+k')x}}{\hat\al(k)\hat
    \al(k')}dkdk',
\end{eqnarray*}
where $\delta\hat\al(k,k'):=\hat\al(0)+\hat \al(k+k')-\hat \al(k)-\hat \al(k')$ and 
\begin{eqnarray*}
&& r_{x}^{(2,1)}= r_x^{(1,2)}= \sum_{x'} \alpha_{x-x'}\left\{\bbE[ (\frak q_x
 -\frak q_{x'}) (\tilde\om*{\frak q})_0]\right\}^2\\
&&
={\cal E}_0^2\int_{\bbT^2}\delta\hat \al(k,k')\frac{e^{2\pi i(k+k')x}}{\om(k)\om(k')}dkdk',
\end{eqnarray*}
$$
r_x^{(2,2)}= 2\left\{\bbE[ 
 (\tilde\om*{\frak q})_0 (\tilde\om*{\frak q})_x]\right\}^2
=2{\cal E}_0^2\delta_{x,0},
$$
\begin{eqnarray*}
&& r_{x}^{(3,1)}= r_x^{(1,3)}= -\sum_{x'} \alpha_{x-x'}\hat\al(0)\left\{\bbE[ (\frak q_x
 -\frak q_{x'}) {\frak q}_0]\right\}^2\\
&&
=-{\cal E}_0^2\int_{\bbT^2}\hat\al(0)\delta\hat \al(k,k')\frac{e^{2\pi i(k+k')x}}{\hat\al(k)\hat\al(k')}dkdk',
\end{eqnarray*}
\begin{eqnarray*}
&&
 r_{x}^{(3,2)}= r_x^{(2,3)}=-2\hat\al(0)\left\{\bbE[  (\tilde\om*{\frak q})_x{\frak q}_0]\right\}^2
\\
&&
=-2{\cal E}_0^2\int_{\bbT^2}\hat\al(0)\frac{e^{2\pi i(k+k')x}}{\om(k)\om(k')}dkdk',
\end{eqnarray*}
and
$$
 r_{x}^{(3,3)}=2\hat\al^2(0)\left\{\bbE[  {\frak q}_x{\frak q}_0]\right\}^2
=2{\cal E}_0^2\int_{\bbT^2}\hat\al^2(0)\frac{e^{2\pi i(k+k')x}}{\hat\al(k)\hat\al(k')}dkdk'.
$$
Therefore, 
we can write \eqref{012204} with
$$
F(k,k'):=\frac{1}{\hat\al(k)\hat
    \al(k')}\left\{-2\hat\al(0) +2\om(k)\om(k')+\delta\hat \al(k,k')\right\}^2.
$$
We have $F(-k,k)=0$. Repeating the argument made in the unpinned case,
this time easier since we do not have to bother about possible
singularities of $F(-k-\eps p,k)$ in the vicinity of $0$ we conclude the
assertion of the proposition
for pinned chains.

\bigskip

\section{Auxiliary results}

\label{auxil}
\subsection{Some computations concerning the scattering kernel}

Directly from \eqref{r} it follows that
 \begin{equation}
 \label{anti-s}
 r(-k,-k')=-r(k,k')
 \end{equation}
 and
  \begin{eqnarray}
  \label{012110}
&&
 r\left(k-\frac p2,k-k'\right) r\left(k+\frac p2,k-k'\right)\nonumber
\\
&&
\\
&&
=16\left[{\frak s}^2\left( k\right)-{\frak s}^2\left( \frac{
      p}{2}\right)\right]{\frak s}^2(k')
\left[{\frak s}^2\left( k+k'\right)-{\frak s}^2\left( \frac{ p}{2}\right)\right].\nonumber
 \end{eqnarray}
From \eqref{R} and  \eqref{anti-s}  we have
\begin{equation}
\label{pm}
R(\pm k,\pm k', \pm p)= R(k,k', p).
\end{equation}
%
%
Equality \eqref{012110} allows us to write the following expansion
\begin{equation}
\label{022110}
R(k,k',p)
= R(k,k')-{\frak s}^2\left( \frac{ p}{2}\right) R_1(k,k')+{\frak s}^4\left( \frac{ p}{2}\right) R_2(k,k').
\end{equation}
Here ${\frak s}(p)$, $R(k,k')$ are $R_1(k,k')$ are given by
\eqref{021701}, \eqref{060411} and \eqref{032110a}, respectively,
 and
$
R_2(k,k')
=8\frak{f}_+(k')
$ (see \eqref{011701}).
Using \eqref{beta} we conclude
\begin{equation}
\label{beta1}
R'(k)=2\pi ({\frak s}(2 k)+{\frak s}(4 k))
\end{equation}
and
\begin{equation}
\label{beta2}
R''(k)=
4\pi^2(4{\frak c}^2(2 k)+{\frak c}(2 k)-2).
\end{equation}
Recall that $R_\eps(k)$ is given by \eqref{022312}.
 Since $R''(0)=12\pi^2>0$, see \eqref{beta2}, we conclude that for any $M>0$
one can find $\eps_0>0$, for which
\begin{equation}
\label{041612}
R_\eps(k)\approx  R(k)+(\eps p)^2,\quad \forall\,k\in \bbT,\,\eps\in(0,\eps_0),\,|p|\le M.
\end{equation}
\begin{lemma}
\label{lm013112}
If $\om(k)\approx |\sin(\pi k)|$ then, for any $M>0$
one can find $\eps_0>0$, for which
\begin{equation}
\label{013112}
\bar \om(k,\eps p)\approx R_\eps^{1/2}(k),\quad k\in\bbT,\,\eps\in(0,\eps_0),\,|p|\le M.
\end{equation}
If on the other hand  $\om(0)>0$
we have
\begin{equation}
\label{013112a}
|\delta_\eps \om(k,p)|\preceq R_\eps^{1/2}(k),\quad k\in\bbT,\eps\in(0,\eps_0),\,|p|\le M.
\end{equation}
\end{lemma}
\subsubsection*{Proof of \eqref{013112}}
Using \eqref{om2} we obtain that 
$$
\bar \om(k,\eps p)\approx \left[\left|\sin\left(\pi\left(k-\frac{\eps
          p}{2}\right)\right)\right|+\left|\sin\left(\pi\left(k+\frac{\eps
          p}{2}\right)\right)\right|\right]
          $$
          for any $k\in\bbT$, $\eps\in(0,1]$, $p\in\bbR$.
Hence, for any $M>0$
one can find $\eps_0>0$, for which
\begin{equation}
\label{011712}
\bar \om(k,\eps p)\approx |\sin(\pi k)|+\eps|p| ,\quad \forall\,k\in\bbT,\,\eps\in(0,\eps_0),\,|p|\le M.
\end{equation}
Estimate \eqref{013112} follows from \eqref{beta} and \eqref{041612}.

\subsubsection*{Proof of \eqref{013112a}}
Note that in case $\om(0)>0$ we have $\om\in C^2(\bbT)$. Since
$\om(k)$ is even we have
$\om'(0)=0$, therefore
\begin{equation}
\label{043112}
|\om'(k)|\preceq |\sin(\pi k)|,\quad k\in\bbT.
\end{equation}
Assume that $p\ge0$. The case $p<0$ can be handled in a similar
fashion. We can write 
\begin{eqnarray}
\label{020611}
&&\delta_\eps\om(k,p)-\om'(k)p=\frac{1}{\eps}\int_{-\eps p/2}^{\eps
  p/2}[\om'(k+h)-\om'(k)]dh\nonumber\\
&&
\\
&&
=\frac{1}{\eps}\int_{0}^{\eps p/2}[\om'(k+h)-\om'(k)]dh+\frac{1}{\eps}\int_{0}^{\eps p/2}[\om'(k-h)-\om'(k)]dh.\nonumber
\end{eqnarray}
The absolute value of the right hand side
of \eqref{020611} equals
\begin{eqnarray}
\label{053112}
&&
\frac{1}{\eps}\left|\int_{0}^{\eps
  p/2}dh\int_0^h[\om''(k+h_1)-\om''(k-h_1)]dh_1\right|
\\
&&
\le \frac{
  p^2\eps}{4}\sup_{|h_1|\le \eps p/2}|\om''(k+h_1)-\om''(k-h_1)|.\nonumber
\end{eqnarray}
From \eqref{043112}--\eqref{053112} it follows that
\begin{equation}
\label{063112}
|\delta_\eps\om(k,p)|\preceq p\left(|\sin(\pi k)|+\eps p\right),\quad k\in \bbT,\,\eps>0,\,p\ge 0.
\end{equation}
Combining this with \eqref{041612} we conclude \eqref{013112a}. \qed


\subsection{Proof of Proposition \ref{prop011601}}

\label{sec14.2}

 From the third equation of \eqref{exp-wigner-eqt-1k} 
\begin{equation}
\label{011401-15}
\bar u_{\eps,-}=\left(\la +\frac{2\ga}{\eps^{\delta}}R_\eps\right)^{-1}\left\{\widehat U_{\eps,-}(0)-\frac{2\bar\om}{\eps^{\delta}}
\bar u_{\eps,+}
-\frac{i\ga R'p}{2\eps^{\delta-1}}(\bar
w_{\eps,-}-\bar w_{\eps})
 +\eps^{3-\delta}\bar r^{(3)}_\eps\right\}.
\end{equation}}
Therefore
\begin{eqnarray}
\label{031401-15}
&&\int_{\bbT}R_\eps\left[\bar\om\left(\la\eps^{\delta} +\ga R_\eps\right)^{-1}\right]^2 |\bar u_{\eps,+}|^2dk
\preceq \int_{\bbT}R_\eps|\bar u_{\eps,-}|^2dk\nonumber\\
&&
+\int_{\bbT}R_\eps\left[\eps^{\delta}\left(\la \eps^{\delta}
  +\ga R_\eps\right)^{-1}\right]^2\left|\widehat
  U_{\eps,-}(0)\right|^2dk+\int_{\bbT}R_\eps\left[\ga\eps R' \left(\la\eps^{\delta}
  +\ga R_\eps\right)^{-1}\right]^2\left|\bar w_{\eps}\right|^2dk\nonumber\\
&&
+\int_{\bbT}R_\eps\left[\eps^3 \left(\la\eps^{\delta}
  +\ga R_\eps\right)^{-1}\right]^2\left|\bar r^{(3)}_\eps\right|^2dk.
\end{eqnarray}
Denote the terms appearing on the right hand side by $J_j$,
$j=1,2,3,4$.
Thanks to \eqref{021601c} we have 
$
J_1\preceq\eps^{\delta-s}.
$
Also, (since $\la\ge\la_0$)
$$
R_\eps\left[\eps^{\delta}\left(\la \eps^{\delta}
  +\ga R_\eps\right)^{-1}\right]^2\preceq R_\eps\eps^{2\delta} \eps^{-\delta}
  \ga^{-1} R_\eps^{-1}\preceq \eps^{\delta-s}
$$
and (since $(R')^2\preceq R$) we have
$$
R_\eps\left[\ga\eps R'\left(\la\eps^{\delta}
  +\ga R_\eps\right)^{-1}\right]^2\preceq R_\eps^2\ga^2\eps^2 \ga^{-2}
R_\eps^{-2}\preceq \eps^2.
$$
From here we get that 
$
J_j\preceq\eps^{\delta-s}$, $j=2,3.$
Finally,
$$
R_\eps\left[\eps^3 \left(\la\eps^{\delta}
  +\ga R_\eps\right)^{-1}\right]^2\le
R_\eps\eps^6\eps^{-\delta}\ga^{-1} R_\eps^{-1}\preceq \eps^{6-\delta-s},
$$
which also yields
$
J_4\preceq\eps^{\delta-s}
$  that finally leads to an estimate 
\begin{equation}
\label{031401-15a}
\int_{\bbT}R_\eps\left[\bar\om\left(\la\eps^{\delta} +\ga R_\eps\right)^{-1}\right]^2 |\bar u_{\eps,+}|^2dk
\preceq \eps^{\delta-s}. 
\end{equation}
To obtain the estimate of ${\cal D}\left(\bar
  w_{\eps}(\la,p)\right)$  it suffices to prove that
${\cal D}\left(\bar u_{\eps,+}(\la,p)\right)\preceq \eps^{\delta-s}$,
which follows, provided we can show that
\begin{equation}
\label{012401-15}
\int_{\bbT}R(k) |\bar u_{\eps,+}(\la,p,k)|^2dk\preceq \eps^{\delta-s}.
\end{equation}
Divide integration in \eqref{012401-15} into two regions: $[|k|\le
\eps^{(\delta-s)/2}]$ and $[|k|\ge \eps^{(\delta-s)/2}]$. In the first
region we  use $R(k)\preceq \eps^{\delta-s}$ and the bound on the $L^2(\bbT)$ norm of  $\bar
u_{\eps,+}(\la,p)$, see \eqref{C-I}. 
Since, in the second region,
$
R_\eps\left[\bar\om\left(\la\eps^{\delta} +\ga
    R_\eps\right)^{-1}\right]^2$ is bounded from below by $ \ga^{-2},
$
in the unpinned case (cf \eqref{013112})
and by
$
\ga^{-2}R_\eps^{-1}
$
in the pinned one we can bound the integral over the region by
$\eps^{\delta-s}$, due to \eqref{031401-15a}. Hence,
\eqref{012401-15} follows.\qed



{\section*{ Acknowledgements.} The authors would like
  to express their gratitude to an anonymous referee for a careful
  reading of the manuscript and remarks that lead to the improvement
  of the manuscript. }

 \end{document}